\documentclass[prd,nofootinbib]{revtex4}
\pdfoutput=1

\usepackage[utf8]{inputenc}
\usepackage[T1]{fontenc}
\usepackage{times}
\usepackage{amsmath}
\usepackage{amssymb}
\usepackage{amsfonts}
\usepackage{amsthm}
\usepackage{mathrsfs}
\usepackage{slashed}
\usepackage{hyperref}
\usepackage{epsfig}                         % roman hinzugefuegt
\usepackage{graphics,graphpap}
\usepackage{graphicx,psfrag}
\usepackage{epstopdf}
\usepackage{color}

\newcommand{\al}{\alpha}
\newcommand{\be}{\beta}
\newcommand{\ga}{\gamma}
\newcommand{\de}{\delta}
\newcommand{\la}{\lambda}
\newcommand{\as}{\alpha_s}
\newcommand{\GeV}{\,\mbox{GeV}}
\newcommand{\MeV}{\,\mbox{MeV}}

\newcommand{\matel}[3]{\langle #1|#2|#3\rangle}

\newcommand{\BKll}{B \to K \ell \ell}
\newcommand{\BKsll}{B \to K^{*} \ell \ell}
\newcommand{\Ecm}{E_{\rm cm}}
\newcommand{\re}{{\rm Re}}
\newcommand{\im}{{\rm Im}}

\newcommand{\Op}{\mathcal O}
\newcommand{\CBBL}{C}
\newcommand{\cc}{{ \eta_{c}}}
\newcommand{\ccm}{\eta_c}
\newcommand{\ccp}{\eta'_c}
\newcommand{\nn}[1]{ \eta_{{\cal B},#1} }
\newcommand{\n}{{ \eta_{\cal B}}}
\newcommand{\cnf}{a_{\rm \fac}}
\newcommand{\qref}{s_{D\!\bar D} }
\newcommand{\qmax}{s_{\rm max}}
\newcommand{\fac}{{\rm fac}}
\newcommand{\cor}{{\rm cor}}
\newcommand{\cut}{{s_{J/\Psi}}}
\newcommand{\HH}{H}
\newcommand{\RR}{\mathbb{R}}
\newcommand{\CC}{\mathbb{C}}
\newcommand{\Disc}{{\rm Disc}}
\newcommand{\dof}{{\rm d.o.f.}}
\newcommand{\hadron}{{\rm QCD}}
\newcommand{\parton}{{\rm pQCD}}
\newcommand{\aver}[1]{\langle #1\rangle}
\newcommand{\onefb}{$1 {\rm fb}^{-1}$}
\newcommand{\threefb}{$3 {\rm fb}^{-1}$}
\newcommand{\Ra}{\Psi(3370)}
\newcommand{\Rb}{\Psi(4040)}
\newcommand{\Rc}{\Psi(4160)}
\newcommand{\Rd}{\Psi(4415)}
\newcommand{\spec}[3]{\!\!\!\phantom{x}^{#1}#2_{#3}}

\newcommand{\Ree}{G(3940)}
\newcommand{\Rf}{Y(4008)}
\newcommand{\Rg}{Y(4260)}
\newcommand{\Rh}{Y(4360)}

\newcommand{\vv}{\kappa}
\def \thl {{\theta_\ell}}
\def \thK {{\theta_{K}}}

\usepackage[all]{xy}
\begin{document}
%\begin{titlepage}
\begin{flushright}\begin{tabular}{l}
Edinburgh/14/10 \\
CP$^3$-Origins-2014-021 DNRF90  \\ 
DIAS-2014-21
\end{tabular}
\end{flushright}

\vskip1.5cm
\begin{center}
  {\Large \bf \boldmath Resonances gone topsy turvy - \\[0.25cm]
 the charm of QCD or new physics in $b \to s \ell^+ \ell^-$?} 
  \vskip1.3cm 
  {\sc James Lyon$^{\,a}$%\footnote{@}  
  \&
    Roman Zwicky$^{\,a}$\footnote{Roman.Zwicky@ed.ac.uk}}
  \vskip0.5cm
  
  $^a$ {\sl Higgs centre for theoretical physics \\
  School of Physics and Astronomy, \\
  University of Edinburgh,   Edinburgh EH9 3JZ, Scotland} \\
 
  \vspace*{1.5mm}
\end{center}

\vskip0.6cm

\begin{abstract}
We investigate the interference pattern of the charm-resonances $\Psi(3370)$, $\Psi(4040)$, $\Psi(4160)$ and $\Psi(4415)$ with the electroweak penguin operator 
$O_9 \propto \bar b \gamma_\al s_L \bar \mu   \gamma^\al \mu$ in 
the branching fraction of $ B^+ \to K^+ \mu \mu$.  
For this purpose we extract the  charm vacuum polarisation via a  standard dispersion relation from  
BESII-data on $e^+e^- \to {\rm hadrons}$. 
In the factorisation approximation (FA) the vacuum polarisation describes the  interference fully non-perturbatively. 
The observed interference pattern  by the LHCb collaboration  
is opposite in sign \emph{and} and significantly enhanced as compared to the FA.
A change of  the FA-result by  a factor of $-2.5$, which correspond to a $350\%$-corrections, results in a reasonable agreement with the data.
This raises the question on the size of  non-factorisable corrections  which are colour enhanced but 
$\al_s/(4 \pi)$-suppressed. 
In the parton picture it is found that the corrections  are of relative size $\simeq -0.5$ 
when averaged over   the open charm-region which is far below $-3.5$ needed to explain the observed effect.
We present combined fits to the BESII- and the LHCb-data, testing for effects beyond the Standard Model (SM)-FA.  
We cannot find any significant evidence 
of the parton estimate being too small due to cancellations between the individual resonances.   
It seems difficult to accommodate the LHCb-result  in the standard treatment of the SM or QCD respectively. 
In the SM the effect can be described in a $q^2$-dependent  (lepton-pair  momentum)  shift of the 
Wilson coefficient combination $C^{\rm eff}_9 + C^{' \rm eff}_9$.  
We devise strategies to investigate the microscopic structure  in future measurements. 
For example a determination of  $C^{\rm eff}_9 - C^{'\rm eff}_9$, in the open charm-region, from $B \to K_{\rm longitudinal}^* \ell \ell$ or $B \to K_0^* \ell \ell$  differing from $C^{\rm eff}_9 + C^{' \rm eff}_9$  implies   the presence of right-handed currents and physics beyond the SM.
We  show that the charm-resonance effects  can accommodate 
the $B \to K^* \ell \ell$-anomalies (e.g. $P_5'$) of the year 2013. 
Hence our findings indicate that the interpretation of the anomaly through a $Z'$-boson, mediating between $bs$ and $\ell\ell$ fields, 
is disfavoured.
More generally our results motivate (re)investigations into $b \to s \bar c c $-physics  
such as the $B \to (\bar cc) K^{(*)}$ decays.
\end{abstract}

\setcounter{footnote}{0}
\renewcommand{\thefootnote}{\arabic{footnote}}

\maketitle

\tableofcontents

%\end{titlepage}

\section{Introduction}

We investigate the pronounced resonance-structure found by the LHCb-collaboration 
\cite{LHCb13_resonances} in  $B^+ \to K^+ \ell \ell $ in the open charm-region. The latter corresponds to high lepton  pair momentum invariant mass $q^2$ (low recoil). 
In the factorisation approximation (FA), where no gluons are exchanged between the charm-loop and the  decaying quarks $b \bar q \to s \bar q$, the charm-resonance contribution is exactly given by the charm vacuum polarisation. The latter can be extracted, to 
the precision allowed by the experiment, from the $e^+e^- \to {\rm hadrons}$ spectrum 
through first principles by a dispersion relation.  

We find that the observed interference pattern has the wrong sign and in addition is more 
pronounced in the data! 
Non-factorisable corrections are assessed, following earlier work in the literature,  
by integrating out the charm quarks and expanding in $1/q^2$ for $q^2 > 14 \GeV^2$.
The relevant contributions (leading to the local resonance structure) are the discontinuities in the amplitude integrated over a 
suitable duality interval.  
We find that the integrated effect amounts to a relative correction of $\sim -0.5$. 
We probe for possible cancelation effects, under the duality integral, by performing 
combined fits of the BESII- and LHCb-data. Cancellation effects, namely varying phases of 
the residues of the resonance poles, are not very pronounced.  We are led to conclude that 
our approach to QCD cannot explain the excess of about a factor $\sim -3.5$  with respect to the FA.

In a second part we devise strategies to unravel the microscopic origin of the effect. 
A promising pathway is to measure the opposite parity combination of Wilson coefficients, that 
enters $\BKll$, in $B_{(s)}  \to K^*(\phi) \ell \ell  $ or $B \to K_0^*\ell \ell$ decays for instance. We also show that 
the effects can explain the $\BKsll$-anomalies, of the year 2013, in the form factor insensitive observable $P_5'$ below the charmonium threshold.

The paper is organised as follows.
In section \ref{sec:BES} we extract the charm vacuum polarisation from BESII-data. 
The FA in $\BKll$ is assed in section \ref{sec:fac} and in section \ref{sec:combined} 
we perform combined fits to the LHCb- and BESII-data. The possible size of non-factorisable corrections are assessed, in some detail, in section \ref{sec:fac-corr}.
Strategies to assess the microscopic origin of the effects are presented in section \ref{sec:strategyandspec}. Further it is shown that the effect can easily accommodate the 
LHCb-anomalies in the low $q^2$-region in $\BKsll$. We end with summary and conclusions in section \ref{sec:discussion}.

\section{Charm vacuum polarisation from  BESII-data}
\label{sec:BES}

We  extract the charm vacuum polarisation $h_c(s)$ from 
the BESII data on $e^+e^- \to \text{hadrons}$. We refer the reader to the textbooks \cite{books} 
for reference of the beginning of this section.
By virtue of the optical theorem the imaginary part of the vacuum polarisation 
is related to the experimentally accessible $R$-function
\begin{equation}
\label{eq:R}
R(s) \equiv \frac{\sigma(e^+e^- \to \text{hadrons})}{\sigma(e^+e^- \to \mu^+\mu^- )} \;,
\end{equation}
 as follows
\begin{equation}
\label{eq:3}
\frac{\pi}{3} R_c(s) =  \frac{1}{2 i} \Disc [h_c](s)    = \im [h_c](s)  \;.
\end{equation}
$\Disc [f](s)\equiv f(s+i0)-f(s-i0) $ denotes the discontinuity of the function $f$ at the point $s$. In the case  at hand  this discontinuity is related to the imaginary part by virtue of the Schwarz reflection principle.
The variable $s \equiv q^2 = \Ecm^2$ denotes  the square of centre of mass energy of the $e^+e^-$-pair.
The meaning of the  superscript $c$ will be clarified in the next section.
The vacuum polarisation $h(s)$ is obtained, through first principles, from the imaginary part (discontinuity) via a once subtracted dispersion relation,
 \begin{equation}
 \label{eq:dispersion}
h_c(s) = h_c(s_0) + \frac{s-s_0}{2  \pi i} P \int_{\cut}^{\infty}  \frac{dt}{ t-s_0 }\frac{\Disc [h_c](t)}{t-s-i0} \;,
\end{equation}
where $s_0 < \cut$ and $\cut   = m_{J/\Psi}^2- \Delta$ where $\Delta > 0$ is sufficiently large such that 
the tail of $J/\Psi$ is covered. In Eq.~\eqref{eq:dispersion} $P$ stands for the Cauchy principal part value. The subtraction is necessary in order to regulate 
the logarithmic ultraviolet UV divergence of the vacuum polarisation.  
The subtraction constant is fixed by $h(s_0)$ (which is real) by perturbative QCD to a  very good approximation 
since it is far away from the $J/\Psi$-resonance.
In summary given arbitrarily precise data on $R$ one can obtain the vacuum polarisation 
to arbitrary precision.  This is a rather fortunate situation with clear potential for future improvement through  
more extensive experimental investigation.

\subsection{Fitting $R_c(s)$ ($3.7 \GeV \leq \sqrt{s} \leq 5 \GeV $) from  BESII-data}

We redo the fit to the BESII-data  ourselves in order to gain control over correlated errors. 
The uncertainty of the BESII-data is around $6.6\%$ and constitutes a significant improvement over previous experiments 
which we do not take into consideration. We refer the reader to the particle data group \cite{PDG} for further reference. 
Below we  describe the separation of the charm part  from the light quark flavours, summarise the experimental uncertainties  
and briefly discuss the fit-model. 
We \emph{stress} that in principle the fit-model is not important as far as this work is concerned since we do \emph{not} aim at extracting resonance parameters such as mass, partial and total widths. Any good fit of the data gives $R_c(s)$ and by \eqref{eq:dispersion} the full vacuum polarisation.
The fit-model is though important as the more realistic it is, the smaller the systematic uncertainty 
through model-bias.

%The $R$-function \eqref{eq:R} is related to the imaginary part of the vacuum polarisation of the electromagnetic current 
%via the optical theorem.  
%As the $e^+e^-$ centre of mass energy increases more and more hadrons get produced.  
We denote by $R_x$, the $R$-function in a world where the photon exclusively  couples to 
the quark flavour(s) $x$. It is our goal to extract $R_c(s)$ since 
this is the quantity that enters into the factorisable charm contribution in decays of the  $\BKll$-type.   
Neglecting interference\footnote{For fixed energy the final states of the 
hadrons produced by light quark currents are in different configurations from the ones of the charm current and hence neglecting interference is justified.}
 between the $u$-, $d$- and $s$- 
current and the $c$-current implies  that $R$ decomposes into 
$R(s) = R_{uds}(s) + R_c(s)$. 
$R_{uds}(s)$ is well-described by perturbative QCD in the region of interest. E.g. at $\Ecm =3.782\GeV$ 
the theory prediction  with uncertainty of about $1\%$ (four loop QCD \cite{KST07}) 
is  consistent with the BESII-data which comes with $~5\%$-uncertainty. Moreover $R_{uds}(s) \simeq 2.16$ 
is quasi-constant over the interval of interest as it is already close to its asymptotic parton-model  value 
of $R_{uds}(s \to \infty) = N_c ( Q_u^2 +Q_d^2 +Q_s^2) = 2$.

The origin of the data and uncertainties necessitate some explanation. 
The original measurement was published in 2002 in \cite{BES02} with statistical and 
systematic uncertainties as given in table III \cite{BES02}.
As explained in \cite{BES02} the common systematic error is $3.3\%$ which we treat as $100\%$-correlated. 
The remaining statistical and systematic errors are treated as uncorrelated.
In 2008 the BESII collaboration fitted the resonance parameters \cite{BES08} 
which in turn led to changes in the initial state radiation correction and results in 
slightly shifted values of the $R$-function c.f. Fig.2 \cite{BES08}. 
This shift is taken into account in our analysis. 

We take the same fit function as BESII with the exception of the continuum background model for which we choose\footnote{The model is chosen such that $R(s) \simeq R_{udsc}$ for $s$ above the resonances and 
the factor $(1-z)$ is the K\"all\'{e}n-function of $\ga^* \to D \bar D$ to the power one and corresponds to an average power  of the various $D \bar D$-final states.
We might improve the matching to pQCD for high $q^2$ in a future version of this work. For the essential points of our analysis this is not of major importance.}
\begin{equation}
\label{eq:Rcon}
R_{\rm con}(s)   =  R_{uds} +  (1-z) (\Delta R_{c} + z a)   \;, \quad \Delta R_{c}  \equiv R_{udsc} - R_{uds}  \;,
\end{equation}
with $z \equiv 4m_D^2/s$,  $R_{uds}=2.16$, $R_{udsc} = 3.6$ and $a$ a fit model parameter.
The values $R_{uds}$ and $R_{udsc} $ correspond to $R(s_1 \equiv (3.73 \GeV)^2 )$ and $R(s_2 \equiv (4.8 \GeV)^2 )$ where predictions of perturbative QCD and BESII experimental data are in impressive agreement.

The transition amplitudes from resonance $r$ to final state $f$, related to the S-matrix as follows $S = 1+ i 2 T$, 
are modelled by a Breit-Wigner ansatz with energy dependent width and interference effects
\begin{equation}
\label{eq:T}
T^{r \to f} (s)  = \frac{m_r \sqrt{\Gamma^{r \to e^+ e^-} \Gamma^{r \to f}(s)}}{s- m_r^2 + i m_r \Gamma_r(s)} e^{i \delta_r}  \;.
\end{equation}
The phase $\delta_r$ is the phase at the momentum of production of the resonance $r$. 
The phase due to $f$ does not need to be written since it cancels out in $R(s)$ on grounds of unitarity of the scattering matrix.
Only single resonances with  quantum numbers of the electromagnetic current ( $J^{\rm PC}  = 1^{--}$)  contribute. 
In the relevant interval, 
\begin{equation}
\label{eq:interval}
\text{fit-interval:} \quad   3.7 \GeV \leq \sqrt{s} \leq 5 \GeV   \;,
\end{equation}
the four $1^{--}$-resonances shown in  table \ref{tab:res}  are fitted for. 
The fit parameters are the interference phases $\delta_r$, the masses $m_r$, the width of the resonance into $e^+e^-$  as well as one normalisation factor for the width into the final states of $D\bar D$-type, based on 
a model by Eichten et al and experimental data, with appropriate thresholds taken into account.  
For further details on the  modelling of  $\Gamma^{r \to f}(s)$, which is rather standard throughout 
the literature,  the reader is referred to the BES-paper \cite{BES08}. 
\begin{table}
\begin{minipage}{0.4\textwidth}
\begin{center}
\begin{tabular}{c | c |c  | c }
     $r$       & $m_r [\GeV]$ & $ \Gamma_r [\MeV] $   & $\spec{2s+1}{L}{J}$  \\  \hline
$J/\Psi$ & $3.097$  &  $0.0934(21)$     &  $\spec{3}{S}{1}$ \\    
$\Psi(2S)$ & $ 3.686$ & $0.337(13)$  &  $\spec{3}{S}{1}$ \\    \hline
$\Ra$  &  $3.771$ & $ 23.3%(5.1)% 
$   &  $\spec{3}{D}{1}$ \\  
$\Rb$  &  $4.039$  &$76.2%(151)% 
$  & $\spec{3}{S}{1}$ \\
$\Rc$  &  $4.192$  & $73.5%(429)% 
$ & $\spec{3}{D}{1}$ \\
$\Rd$  &  $4.415$  & $78.5%(571)%
$  & $\spec{3}{S}{1}$
\end{tabular}
\caption{\small $J^{\rm PC}  = 1^{--}$ charmonium resonances.  The first two resonances are narrow and the uncertainties in the masses are negligible \cite{PDG}. The last four resonances are above the $D\bar D$-threshold and as a consequence the width is much larger. We have taken 
our central fit values from table \ref{tab:BESfit}, where more details on the fit can be found. The uncertainty in the widths is considerable.}
% used in the fit \eqref{eq:Rres}  \newline \newline
%$\phantom{xxxxxx}$   $\phantom{x}$ 
\label{tab:res}
\end{center}
\end{minipage}
\qquad
\begin{minipage}{0.4\textwidth}
\begin{center}
\begin{tabular}{c | c | c }
      $r$      & $m_r [\GeV]$ & $\Gamma_r[\MeV] $  \\ \hline
$\Ree$  &  $3.943(21)$ &  $52(11)$ \\  
$\Rf$  &  $4.008^{(121)}_{(49)}$  &  $226(97)$ \\
$\Rg$  &  $4.263(5)$  &  $108(14)$ \\
$\Rh$  &  $4.353(11)$  &  $96(42)$
\end{tabular}
\caption{\small $J^{\rm PC}  = 1^{--}$ exotic (non-charmonium) resonances currently \emph{not} yet used in the fit.  Relevant comments in the main text.}
\label{tab:res2}
\end{center}
\end{minipage}
\end{table}
%Note that the energy dependence of $\Gamma^{r \to f}(s)$ results in an energy dependent width $\Gamma_r(s)$.
The fit function is given by 
\begin{equation}
\label{eq:Rfit}
R_{\rm fit}(s) = R_{\rm res}(s) + R_{\rm con}(s)
\end{equation}
with $R_{\rm con}$ as in  \eqref{eq:Rcon} and the resonance part as given by
\begin{equation}
\label{eq:Rres}
 R_{\rm res}(s) = \frac{9}{\al^2} \sum_f | \sum_r T^{r \to f} (s) |^2  \;.
\end{equation}
The factor $9/\al^2$ comes from the normalisation $\sigma(e^+e^- \to \mu^+\mu^- ) = 4 \pi \al^2 /(3s)$ where 
$\al$ is the QED fine structure constant.  
Since only relative phases are observable the first phase 
  $\delta_{\Ra} \equiv 0$ is set to zero by convention.
This amounts to a total number of $(4\times 4 -1)_{\rm res} +1_{\rm con} = 16$ fit parameters. 
We perform a $\chi^2$ minimisation and obtain a chi squared per degree of freedom (\dof) $\nu = 78- 16 -1 = 61$  of 
 \begin{equation}
 \label{eq:chi2BES}
 \chi^2/{\dof}|_{\rm BESII-data} = 1.015
 \end{equation}
  which corresponds to a $p$-value of $44\%$ and is  close to $\chi^2/{\dof}  = 1.08$ \cite{BES08} as should be the case since
 we employ the same data and a quasi identical model. The fit is shown in Fig.~\ref{fig:h} (top) 
 and the fit parameters are given in table \ref{tab:BESfit} in appendix \ref{app:BESII}. In agreement with  \cite{BES08} we observe that  $\chi^2/{\dof} \simeq 1.35$ when the interference phases $\delta_r$ are omitted from the ansatz \eqref{eq:T}. 

To this end let us comment on the relevance of  exotic charmonium resonances 
discovered throughout the last decade.  
The ones of interest for  our purposes ($1^{--}$ states that located in the fit-interval) are  are listed in table \ref{tab:res2} with numbers taken from the review paper 
\cite{HQium-review}.\footnote{One could also include the $X(4630)$ and  $Y(4660)$ \cite{HQium-review} which are just $\sim 150 \MeV$ below the  
  kinematic endpoint $\qmax \equiv m_B - m_K \simeq 4.8 \GeV$.}

\begin{figure}[h]
 \includegraphics[scale=0.37]{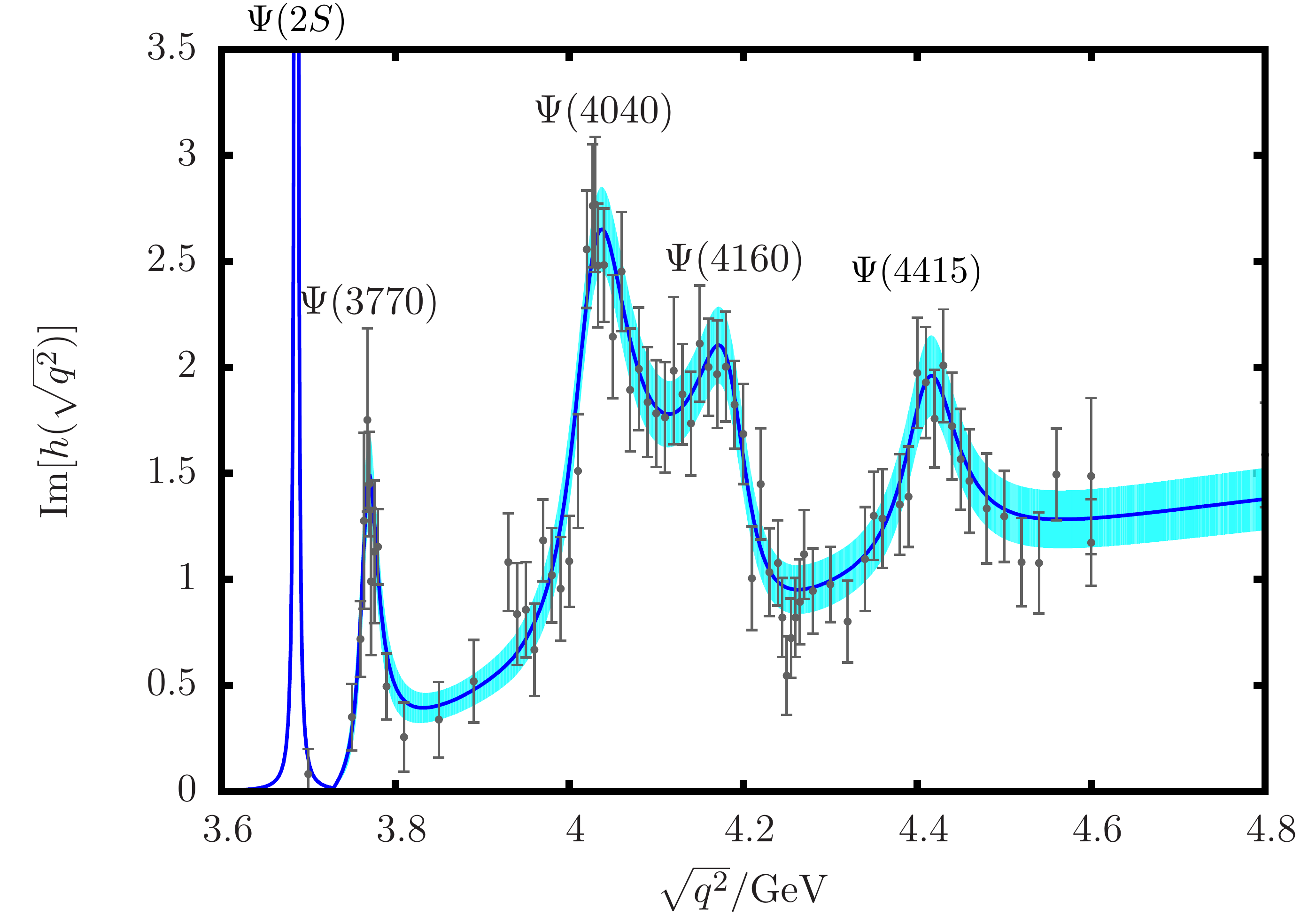} \\ 
 \includegraphics[scale=0.405]{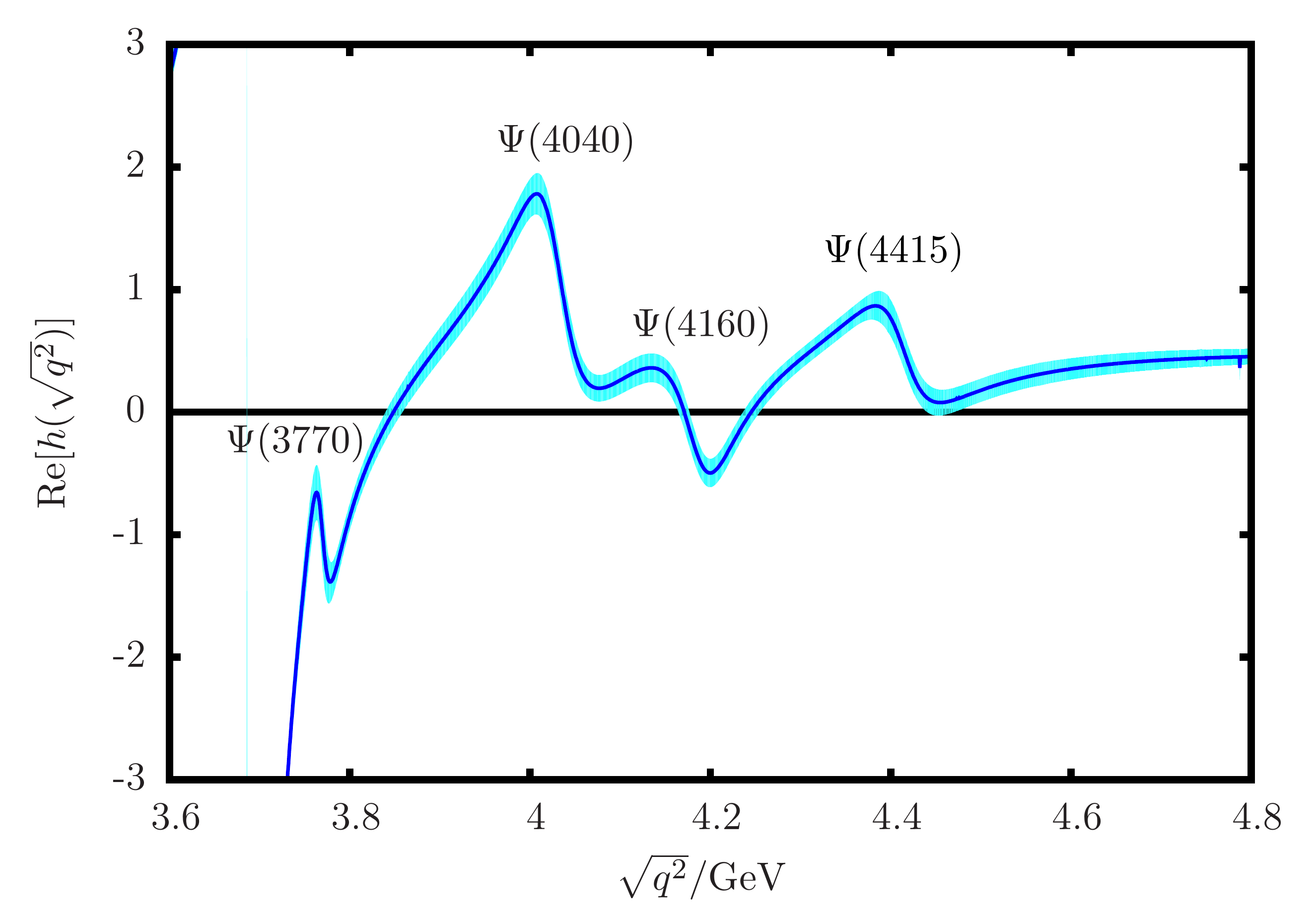}
 \caption{\small (top) Imaginary part of of vacuum polarisation fitted to BESII data. 
In the plot we show the BESII error bars with systematic and statistical uncertainty added in quadrature.  
The $1\sigma$-error band is shown in cyan. (bottom) Real part of the vacuum polarisation obtained 
from \eqref{eq:dispersion} with error band as for the imaginary part.}
 \label{fig:h}
 \end{figure}

%Currently there is no need since the low $\chi^2$ \eqref{eq:chi2BES}
From the viewpoint of the dispersion relation \eqref{eq:dispersion}, it is immaterial, whether the hadronic model is accurate as long as the fit is good which is measured by 
the low $\chi^2$ \eqref{eq:chi2BES}. 
It is conceivable that with more data the inclusion of these states would improve the fit.\footnote{Possibly the $G(3940)$ which is  narrow and known to decay into $D\bar D$ could be included.
 In fact, being biased by the knowledge, one might almost say that they are visible in the BESII-spectrum c.f. 
 Fig.~\ref{fig:h}(top) as well as in the LHCb data shown in Fig.~\ref{fig:rate}. Earlier data  does not seem to indicate a significant raise at $E = 3940 \MeV$ (c.f. plots in \cite{PDG})}
Another  possible  future improvement would be to extend the Breit-Wigner model into 
a K-matrix formalism.

% Second since they are non-charmonium states the interference, as previously noted, is presumably negligible since the decay products are in different configurations.
% Similar things can be said about the $Y(4260)$.

\subsection{Assembling $R_c(s)$ and the vacuum polarisation}
\label{sec:assemble}

The function $R_c(s)$ which we fitted in the interval \eqref{eq:interval} has to be completed below and above 
the fit-interbal  in order to obtain  $\re [h_c]$ through  \eqref{eq:dispersion}.
Fortunately this is no problem e.g. \cite{KS96a}. Below the interval $R_c$ is well-approximated by a 
Breit-Wigner ansatz 
\begin{equation}
\label{eq:narrow}
R_{\text{narrow}}(s) = - \frac{9}{\al^2}  {\rm Im}\Big[ \sum_{r \in \{ J/\Psi,\Psi(2S)\}} 
\frac{m_r \Gamma^{r \to \ell^+\ell^-} }{s-m_r^2 +i m_r\Gamma_r} 
\Big] \;,
\end{equation}
without interference effects since  the $J/\Psi$ and $\Psi(2S)$ are narrow and sufficiently far apart from each other.  
It is noted that \eqref{eq:narrow} relates 
to \eqref{eq:T} through the optical theorem  $\im[T] = T T^\dagger$ when $f = \ell\ell$ and $\delta_r \to 0$.
Above the fit-interval perturbative QCD provides an excellent approximation. Schwinger's $O(\al_s)$-interpolation 
result \cite{Schwinger}, for $s > 4 m_c^2$, reads
\begin{eqnarray}
\label{eq:hPT}
\im[h_c](s)  &=&  \im[h_c^{(0)}](s) + \al_s  \im[h_c^{(1)}](s)  =\frac{2 \pi}{9} (3 -v(s)^2) |v(s)| \Big(1 + 
   \frac{4}{3} \al_s \big(\frac{\pi}{2 v(s)} - \left(\frac{3}{4} + \frac{v(s)}{4}\right) \left(\frac{\pi}{2} - \frac{3}{4 \pi}\right) \big) \Big) \;, \nonumber
\end{eqnarray}
where $v(s) \equiv \sqrt{1 - 4m_c^2/s}$ is proportional to the charm quark momentum in the centre of mass frame of the lepton pair.
The well-known one loop result for real and imaginary  part $h^{(0)}$ is given in Eq.~\eqref{eq:h0} in the appendix.
%For completeness we list:
%\begin{equation}
%\label{eq:Rc}
%R_c(s) = \left\{
%\begin{array}{ll}  R_{\text{narrow}}  & s <  s_1   \\[0.2cm]
% R_{\rm fit} - R_{uds}   &   s_1 < s < s_2 \\[0.2cm]
%R_{\rm pQCD} &    s_2 < s
%\end{array} \right.  
   %   R_{J/\Psi}(s)  + R_{\Psi(2S)} + \tilde R_c(s)  + R_{\rm pert}
%\end{equation}
%where $s_{1,2}$ have been defined above. The real part of the vacuum polarisation, obtained through \eqref{eq:dispersion} and \eqref{eq:Rc} is plotted in Fig.~\ref{fig:h} (bottom).

\section{Factorisation gone topsy turvy}
\label{sec:fac}

\subsection{Effective Hamiltonian}

In the SM the relevant effective Hamiltonian  for $b \to s \ell^+\ell^-$ transitions reads
\begin{equation}
\mathcal H_{\mathrm{eff}} = \frac{G_F}{\sqrt 2}\left( \sum_{i=1}^2 
(\lambda_u C_i \Op_i^u + \lambda_c C_i \mathcal  \Op_i^c )       -\lambda_t \sum_{i=3}^{10} C_i  \Op_i \right)  \;, \qquad 
\lambda_i \equiv V_{is}^*V_{ib} \;,
\label{eq:effective-ew-hamiltonian}
\end{equation}
where the Wilson coefficients $C_i(\mu)$ and the operators $O_i(\mu)$ carry a dependence on the factorisation scale $\mu$ separating the UV from the infrared (IR) physics.
The $b \to s$ unitarity relation reads $\lambda_u + \lambda_c + \lambda_t = 0$. 
For the discussion in this paper we use the basis\cite{Buchalla:1995vs}\footnote{
The sign convention of $\Op_{7,8}$ 
corresponds to a covariant derivative $D_\mu = \partial_\mu - i Q  e A_\mu - i g_s A_\mu$ and  interaction vertex $+i( Qe + g_s\frac{\lambda^a}{2}) \gamma^\mu$.}
\begin{align}
\label{eq:SMbasis}
\mathcal O_1^q &= (\bar s_i q_j)_{V-A}(\bar q_j b_i)_{V-A} &
\mathcal O_2^q &= (\bar s_i q_i)_{V-A}(\bar q_j b_j)_{V-A} \nonumber \\
\mathcal O_3 &= (\bar s_i b_i)_{V-A} \sum_q (\bar q_j q_j)_{V-A} &
\mathcal O_4 &= (\bar s_i b_j)_{V-A} \sum_q (\bar q_j q_i)_{V-A} \nonumber \\
\mathcal O_5 &= (\bar s_i b_i)_{V-A} \sum_q (\bar q_j q_j)_{V+A} &
\mathcal O_6 &= (\bar s_i b_j)_{V-A} \sum_q (\bar q_j q_i)_{V+A} \nonumber \\
\mathcal O_7 &= -\frac{e m_b}{8\pi^2}\bar s \sigma\cdot F (1+\gamma_5)b &
\mathcal O_8 &= -\frac{g_s m_b}{8\pi^2}\bar s \sigma\cdot G (1+\gamma_5) b \nonumber \\
\mathcal O_9 &= \frac{\alpha}{2\pi}(\bar \ell\gamma^\mu \ell)(\bar s \gamma_\mu (1-\gamma_5) b) &
\mathcal O_{10} &= \frac{\alpha}{2\pi}(\bar \ell\gamma^\mu\gamma_5 \ell)(\bar s \gamma_\mu (1-\gamma_5) b) \;, & 
\end{align}
where $i,j$ are colour indices, $(\bar s b)_{V\pm A}=\bar s \gamma^\mu (1\pm\gamma_5) b$,  $e = \sqrt{4 \pi \alpha} > 0 $  and $G_F$ is the Fermi constant. 
%The ingredients to obtain the Wilson coefficients 
%$C_i$ at the scale $\mu$ of order $m_b$ can be found e.g. in 
%\cite{Buchalla:1995vs,Bobeth:1999mk,Gambino:2003zm}.

\begin{figure}[h]
 \includegraphics[scale=0.8]{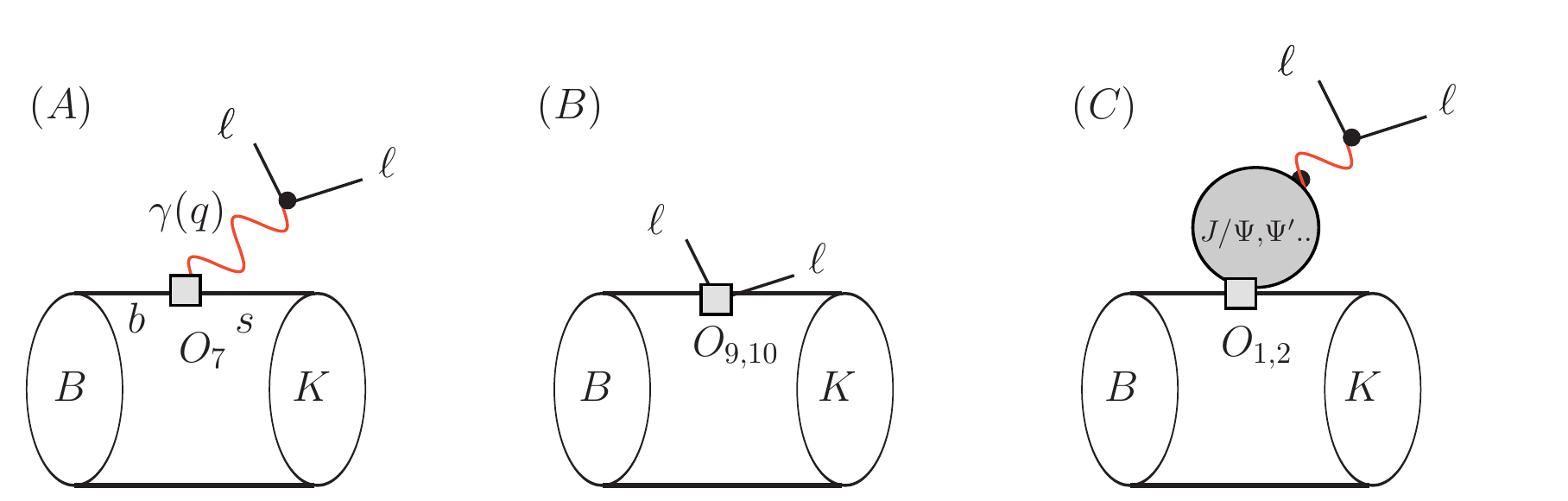}
\caption{\small  Numerically leading contributions to the decay rate of $\BKll$ in the high $q^2$-region.
(a) and (b) $O_7$ and $O_{9,10}$ short distance contributions.  These contributions are proportional to the local (short distance) form factors. (c) long distance charm-loop contribution which 
in (naive) factorisation is proportional to the same form factor times the charm vacuum polarisation $h_c(q^2)$. 
The charm bubble itself is the full non-perturbative vacuum polarisation since it is extracted directly from the data.}
 \label{fig:fac}
 \end{figure}

At high $q^2$, by which we mean above the narrow charmonium resonances, 
the numerically most relevant contributions to $\BKll$ are given by the form factor contributions 
$\Op_7$ and  $\Op_{9,10}$ (c.f. Fig.~\ref{fig:fac}AB as well as the tree-level four quark operators 
$\Op^c_{1,2}$ which have sizeable Wilson coefficients.) In this section we employ the (naive)\footnote{The term naive refers to the fact that in this approximation the scale dependence of the Wilson coefficients $C_i$ 
is not compensated by the corresponding scale dependence of the matrix elements, a point to be discussed in the forthcoming section.} factorisation  approximation (FA) for which,
\begin{equation}
\matel{K}{C_1 \Op_1^c + C_2 \Op_2^c}{B}|_{\rm FA} \propto   ( C_1 + C_2/3 ) f^{B \to K}_+(q^2) h_c(q^2) \;,
\end{equation}
the matrix element factorises into the charm vacuum polarisation $h_c$ times the short distance form factor as defined in Eq.~\eqref{eq:ff}. This contribution has got the same form factor dependence as $C_9$ 
and can therefore be absorbed into an effective Wilson coefficient $C_9^{\mathrm{eff}}$ \eqref{eq:Ceff} and \eqref{eq:Y}. 
The combination $C_1 + C_2/3$ is known as the ``colour suppressed" combination of Wilson coefficients because of a substantial cancellation of the two Wilson coefficients 
(c.f. appendix \ref{app:colour}).
This point will be addressed when we 
discuss the estimate of the ${\cal O}(\al_s)$-corrections.

\subsection{SM-$\BKll$ in  factorisation} 

\begin{figure}[h]
 \includegraphics[scale=0.5]{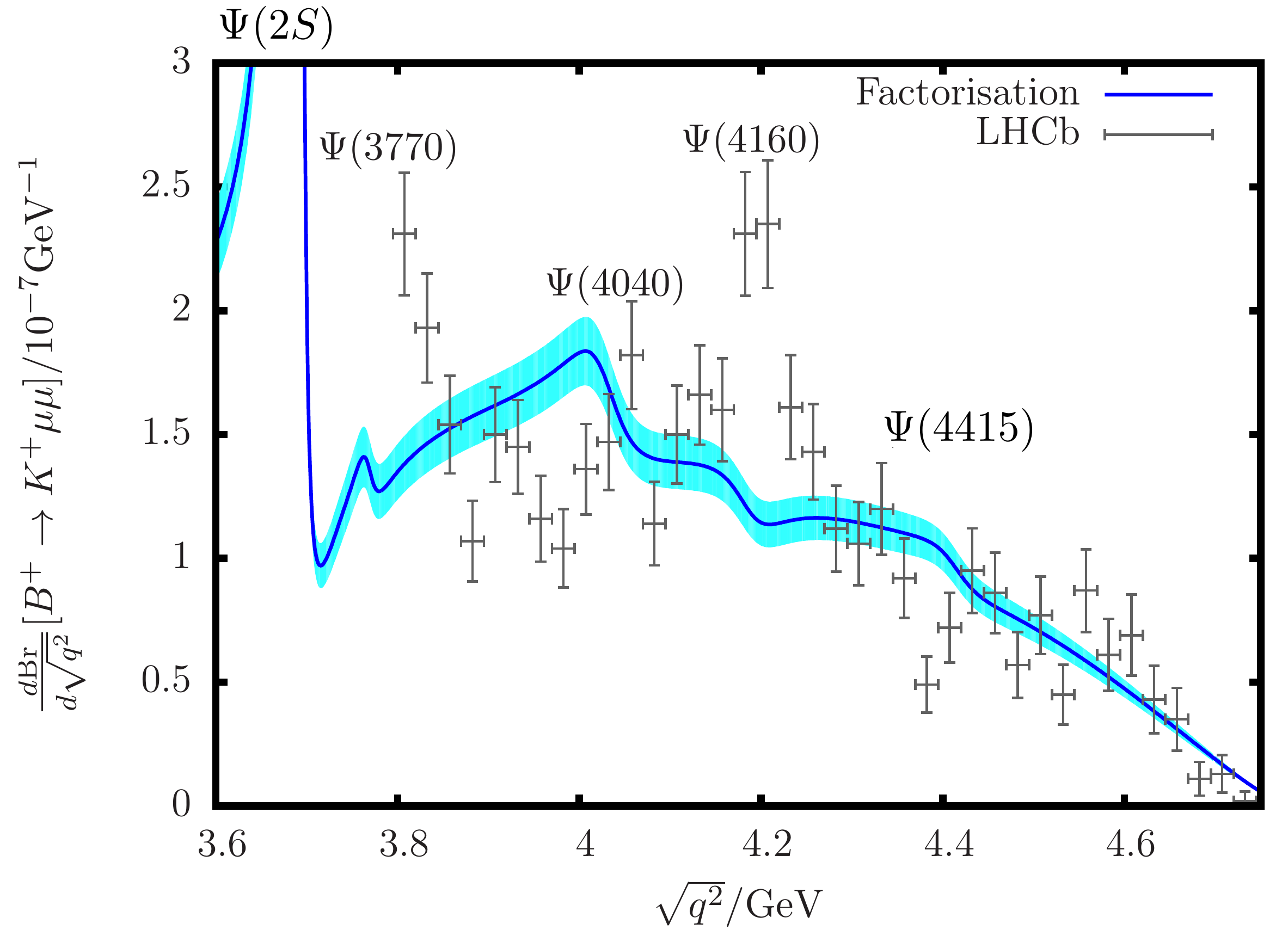}
\caption{\small $\BKll$ rate for high $E \equiv \sqrt{q^2} $ just above the $\Psi(3770)$-resonance up to the kinematic 
endpoint. The $40$ LHCb bins   \cite{LHCb13_resonances,merci}  are shown with grey crosses. 
The solid blue line corresponds 
to our SM prediction using FA  (the non-factorisable corrections are discussed in chapter \ref{sec:fac-corr}). 
The cyan band is the theory error band.
The mismatch between FA and the data  is apparent to the eye.}
\label{fig:rate}
 \end{figure}

Our SM prediction with lattice form factors \cite{Lattice13} (c.f.  appendix \ref{app:input} 
for more details), for the $\BKll$-rate are shown in Fig.~\ref{fig:rate} against the LHCb data  \cite{LHCb13_resonances,merci}. 
It is apparent to the eye that the resonance effects, in (naive) factorisation, turn 
out to have the wrong sign! Not only that but they also seem more pronounced in the data 
which will be reflected in the fits to be described below.

\section{Combined fits to BESII and LHCb data in and beyond factorisation} 
\label{sec:combined}

Before addressing the relevant issue of corrections to 
the SM-FA in section \ref{sec:fac-corr},
we present a series combined fits to the  BESII and LHCb-data. 
We first describe the fit models 
before commenting on the results towards the end of the section. 
The number of fit parameters and the number of \dof, denoted by $\nu$,  are given in  brackets below. 
We take $78$ BESII data points and $39$ LHCb bins, excluding the last bin which has a negative entry, amounting  to a total 
of $117$ data points.

\begin{itemize}
\item [a)] {\bf Normalisation of the rate},  ($17 = 1_{\n} +16_{\rm res}$  fit-parameter $\n$,  $\nu = 117-17-1=99$)   
\\[0.1cm]
In the FA  the normalisation of the
rate is given by the form factors $f_{+,T}(q^2)$. Since the latter are closely related in the high $q^2$-region by Isgur-Wise relation this amounts effectively to an overall normalisation. 
To be precise we parameterise the pre-factor, inserted into \eqref{eq:dGdq2} with $m_l =0$ for the sake of illustration, as follows
\begin{equation}
\label{eq:fp1}
\frac{d\Gamma}{dq^2}^{B\to K \ell^+\ell^-}  \propto  \quad  \n (|\HH^V|^2 + |\HH^A|^2) \;,
\end{equation}
where $V$ and $A$ refer to the lepton polarisation.

\item [b)] {\bf Prefactor of $h_c(q^2)$},  ($18= 2_{\n,\cc} + 16_{\rm res}$ fit parameters, $\nu = 117-18-1= 98$) \\[0.1cm]
In addition to the normalisation, we fit for a scale factor $\cc$ in front 
 of the factorisable charm-loop $h_c(q^2)$. More precisely:
\begin{eqnarray}
\label{eq:hV}
& & \HH^V = C_9^{\rm eff} \frac{(m_B+m_K)}{2m_b} f_+(q^2) + C_7^{\rm eff} f_T(q^2)  \;, \nonumber \\
& & C_9^{\rm eff} = (C_9 +   \cc \cnf h_c(q^2) + ... )
\end{eqnarray}
where $C_9(\mu) \simeq 4$, $C_7^{\rm eff}(\mu) \simeq -0.3$, $\cnf(\mu) \simeq 0.6$ at 
$\mu \simeq m_b$ and $h_c(q^2)$ is shown in Fig.~\ref{fig:h}.  
The dots stand for quark loops of other flavours.
\end{itemize}

In a next step we probe for non-factorisable corrections by letting the fit residues of the LHCb data take on arbitrary real (fit-c) and complex (fit-d) numbers.  
We would like to emphasise that in addition to non-factorisable effects new  operators with $J^{PC}[\bar c \Gamma c ] = 1^{--}$, other than the vector current,  can also lead to such effects. 
More discussion can be found later on.

For the charm vacuum polarisation the discontinuity $\Disc [h_c]$ is necessarily 
positive Eq.~(\ref{eq:Rres},\ref{eq:3}) and its relation to physical quantities is given \eqref{eq:T}. Hence we can test for 
physics beyond SM FA by the following replacement 
\begin{equation}
\label{eq:rhor}
  | \sum_r T^{r \to f} (s) |^2 \to ( \sum_r  \rho_r T^{r \to f }(s)   )  ( \sum_r T^{r \to f }(s)   )^*    \;.
\end{equation}
The scale factor $\rho_r$ roughly corresponds to ${\cal A}( B \to K \Psi )/f^{B\to K}_+(q^2)$ and replaces ${\cal A}( \Psi \to \ell \ell)$ in \eqref{eq:T}.

For the fits c) and d) we are not going to put any background model to the LHCb-fit since with the current  
precision of the LHCb data it seems difficult to crosscheck for the correctness of any model.  
The background is essentially zero at the $\bar D D$-threshold and is expected to raise smoothly  with kinks at the thresholds of various 
$D\bar D$-thresholds  (with the two $D$'s being any of $D ,D^*,D_s,D^*,D_1,\dots $)  
into the region where perturbation theory becomes accurate. In fact this is the essence behind the model ansatz \eqref{eq:Rcon}.
The branching fraction has just got the opposite behaviour to the background and 
this is the reason why  it seems difficult to extract the background from the data. 
More data could, of course, improve the situation.

\begin{itemize}
\item [c)] 
{\bf  Variable residues $\rho_r \in \RR$},  ($22 = 1_{\n} + 5_{\rho_r} + 16_{\rm res}$ fit parameters, $\nu = 117-32-1 = 94$)   \\[0.1cm]
 We choose to keep $\n \equiv 1$ and parameterise $\rho_{\Psi(2S)}$ instead which is an equivalent procedure.
The five parameters $\rho_r$ are constrained to be real.

\item [d)] {\bf Variable residues $\rho_r \in \CC$},  ($27= 1_{\n} + 10_{\rho_r} + 16_{\rm res}$ fit parameters, $\nu = 117-27-1=89$) \\[0.1cm]
Idem but with $\rho_r \in \CC$ allowing for dynamical phases, therefore introducing 5 new fit parameters.
%It should be added that unlike in the FA the phases due to scattering of $r \to f$ are not going to cancel 
%and one should in principle introduced and $f$-dependence in $\rho_r^f$. So effectively the fit assumes universality in those phases 
%which might well be a good assumption.
\end{itemize}

\begin{figure}[h]
 \includegraphics[scale=0.6]{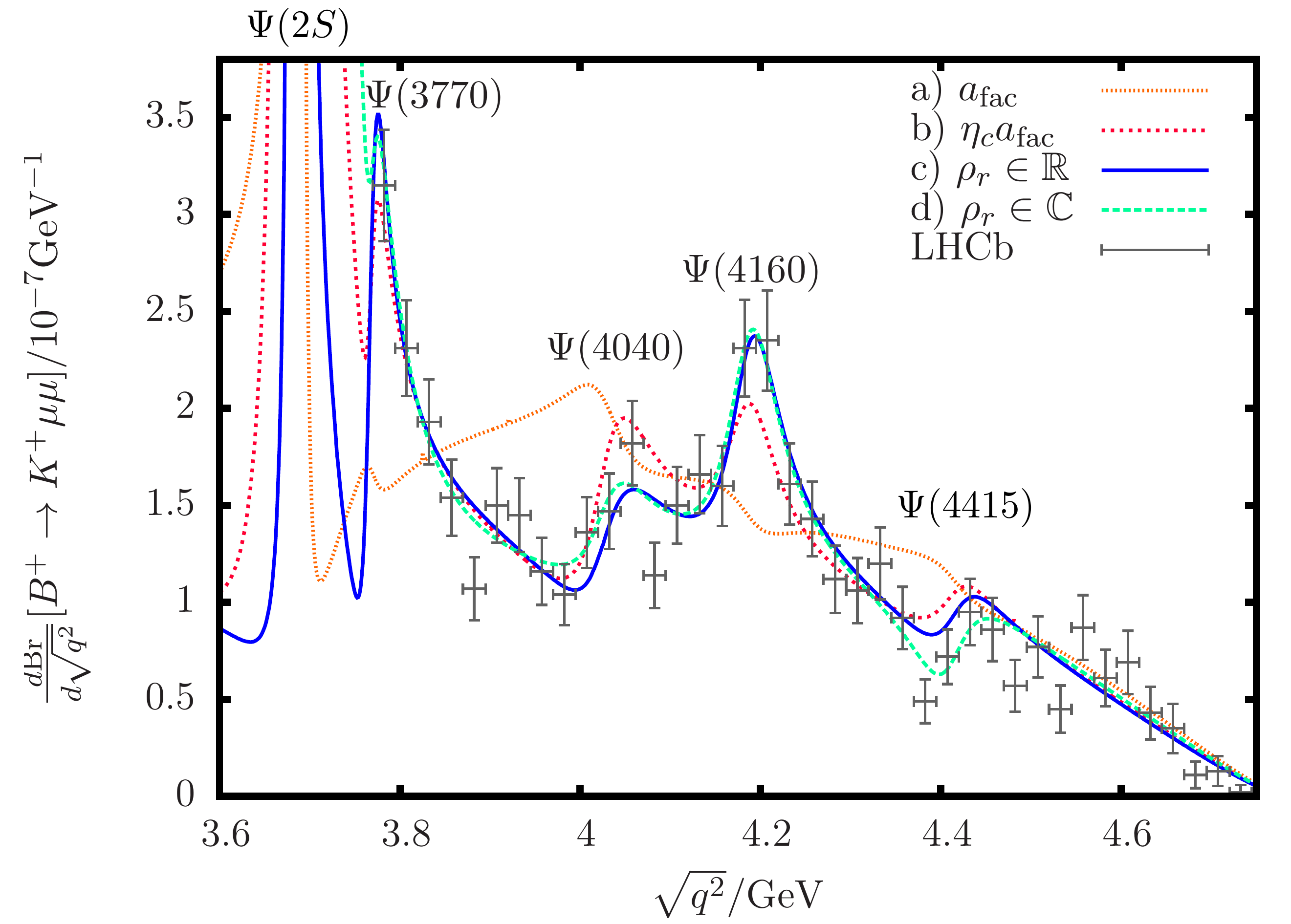}
\caption{\small LHCb-data \cite{LHCb13_resonances,merci}  (grey crosses) versus the fits b) c) d) described in the main text.}
\label{fig:rate2}
 \end{figure}

The fits are shown in Fig.~\ref{fig:rate2} and the values are given in table  \ref{tab:fit1}. 
Some more detail on the fit-procedure  is described 
in appendix \ref{app:combined}. Below we comment on each of the fits.

\begin{itemize}
\item[Fit-a):]
\emph{If}  the FA was a good approximation then the fit on the first line indicates that that the SM 
is excluded  by a 
$p$-value of $\simeq 10^{-30}$.  This is much more significant than $5 \sigma$ (see appendix \ref{app:combined} for a refined remark). 
%It is noteworthy that the best fit results in a shift of the normalisation of $26\%$ which corresponds to roughly $13\%$ downwards shift 
%of the lattice form factors which is a bit more than the corresponding uncertainty prediction 
%in \cite{Lattice13}.
\item[Fit-b):] Allowing for the charm-resonance prefactor $\cc$ to vary gives a (reasonable) $\chi^2$ per degree of freedom with a $p$-value of about $2.1\%$. 
Most noticeably  $\cc = -2.55$ is rather large and negative. The nominal value of non-factorisable corrections correspond to a shift 
of $\Delta \cc \simeq - 0.5$ and hence   $\cc = -2.55$ would indicate an effect which is seven times larger. This statement is to be refined 
under fits c) and d) and the discussion in the next section.
\item[Fit-c,d):]  
It is noticeable that there is no uniformity in the residues which in principle is  a sign for contributions beyond the SM-FA. 
Yet there are two important points we would like to make,
 First the  $\chi^2/$\dof $\simeq 1.17$ and $1.12$  cannot be seen as a drastic 
improvement over $\chi^2/\dof_{\rm fit\; b)} \simeq 1.33 $ and hence it is not clear how much one can read into these fits. 
Second  the residues of $\Psi(2S)$ and $\Ra$ cannot be taken at face value since they essentially have opposite magnitude 
in fits c) and d). Yet, as can be inferred from plots in Fig.~\ref{fig:rate2}, the curves are hardly distinguishable in the relevant region. 
This is an issue that could be improved with further data points below $\sqrt{q^2} = 3.770 \GeV$.  
On average the last three residues reflect the $\n = -2.55$ shift seen in fit-b).\footnote{It is noticeable that the residues of the $\spec{3}{D}{1}$ in fit-c) 
are somewhat  larger than $\spec{3}{S}{1}$  which might be a hint towards the underlying physics driving this effect. This sharpens the demand for more data points in order to resolve the ambiguity of the first two residues between fits c) and d).}

\end{itemize}

  \begin{table}[h]
\begin{center}
\begin{tabular}{c | c   c ||  lllll     ||  r  | r | r | r  }
Fit & $\n$ & $\cc$ & $\rho_{\Psi(2S)}$ & $\rho_{\Ra}$ & $\rho_{\Rb}$ & $\rho_{\Rc}$ & $\rho_{\Rd}$ &   $\chi^2/$\dof  
 & \dof & pts  &  $p$-\text{value}  \\ \hline 
$a)$ &  $0.98$ & $\equiv1$  & $\equiv1$ & $\equiv1$ & $\equiv1$ & $\equiv1$ & $\equiv1$   &  $3.59$ &  $99$ & $117$  & $\simeq 10^{-30}$  \\ \hline
$b)$ & $1.08$ & -$2.55$ & $\equiv1$ & $\equiv1$ & $\equiv1$ & $\equiv1$ & $\equiv1$   &  $1.334$ &  $98$& $117$  & $1.5\%$  \\ 
 %       & "          & $\equiv 1$ & -$2.8$ & -$2.8$ & -$2.8$ & -$2.8$ & -$2.8$ & " & " &" &" &  "  \\
\hline 
$c)$ & $0.81$ & $\equiv  1$ & -$1.3$ & -$7.2$ & -$1.9$ & -$4.6$ & -$3.0$   & $1.169$ &  $94$ & $117$ & $12\%$        \\ \hline
$d)$ & $1.06$ & $\equiv 1$ &   \phantom{-}$3.8$-$5.1i$  & -$0.1$-$2.3i$  & -$0.5$-$1.2i$  &  -$3.0$-$3.1i$ &  -$4.5$+$2.3i$ 
   &  $1.124$ &  $89$ & $117$ & $20\%$    \\ 
  &  &  &   $6.4e^{-i 53.3^\circ }$  & $2.0e^{-i 92^\circ }$  & $1.3e^{-i 111^\circ }$  & $4.3e^{-i 135^\circ }$  & 
  $5.1e^{i 153^\circ }$    &  &  &  & 
  \end{tabular}
\caption{\small  Combined fit to BESII and LHCb-data. The parameter $\n$ \eqref{eq:fp1} is an overall normalisation factor,  
\eqref{eq:hV} is the pre coefficient of $h_c(q^2)$ and  the meaning of $\rho_{\Psi}$ is given in \eqref{eq:rhor}.
The prediction for the  SM-FA  is $(\n,\cc,\rho_\Psi) = (1,1,1)$. For fit-d) we have given the residues in cartesian polar complex coordinates. The background-model fit-parameter  $a$ \eqref{eq:Rcon} for fits a) to d) is given by 
 $a = (2.886  ,2.655 ,3.100 ,3.056   )$ respectively. As explained in the text the background model is not applied to fits c) and d).}
\label{tab:fit1}
\end{center}
\end{table}

\section{Discussion on non-factorisable corrections}
\label{sec:fac-corr}

The size of non-factorisable corrections  in $b \to s \bar cc $-transitions 
is a recurring question since the latter  are not colour suppressed (c.f. appendix \ref{app:colour} for a brief discussion)  as opposed to the factorisable corrections. 
This raises the question of whether or not the non-factorisable contribution,  being $\al_s$ suppressed, 
could dominate as a result of the colour enhancement (or colour non-suppression). 
Our investigation indicates that this is unlikely to be the case for $\BKll$  for $q^2 > \qref$. 

The section is organised as follows.  First the sizeable corrections are identified in subsection \ref{sec:charm_out}. 
In subsection \ref{sec:dispersion}  the topic under investigation is elaborated on from the viewpoint of a dispersion relation, (non)-positivity  and Breit-Wigner resonances.
Finally the size of the correction are estimated through the partonic picture and through the actual data 
in subsections \ref{sec:partonic} and \ref{sec:data-fit}.
This section consists  of a lengthy, but important, chain of arguments. 
The casual reader might want to directly pass to  the final message in subsection \ref{sec:summary}.

\subsection{Integrating out the charm quarks}
\label{sec:charm_out}

To go beyond the FA, in the parton picture, one gluon exchanges between the charm-loop and the $B \to K$-transition need to assessed. 
This is a difficult task in principle. 
In the kinematic situation   $q^2 $ constitutes, fortunately,   a large scale that can be taken advantage of by integrating out the virtual charm  quarks in the loop. 
Using the external field method $ \bar b D^n s$-operators, in increasing dimension ($3 + n)$, are generated. The symbol  $D^n$  represents  $n$ covariant derivatives. 
This approach was  suggested and investigated  in \cite{GP04} within heavy quark effective theory. 
An analysis in QCD (without expanding in the $m_b$-mass)  including modelling of duality violation effects was given in \cite{BBF11}. This framework has become known as the 
``high-$q^2$ operator product expansion (OPE)".
The term OPE is a bit derived in the sense above since strictly speaking an OPE is a short distance expansion whereas the resonance region corresponds to a regime where hadrons propagate over long distances.  
Hence, unlike in the previous section we can  not hope to resolve the corrections locally in  $q^2$ (as shown in the plot in Fig.~\ref{fig:rate}).  Yet one can get an estimate of the 
effect on the helicity amplitudes 
by integrating over suitable, to be made more precise, duality intervals.  
It is this quantity  that we  compare to the FA integrated over the same interval.

\begin{figure}[h]
\begin{center}
 \includegraphics[scale=1.0]{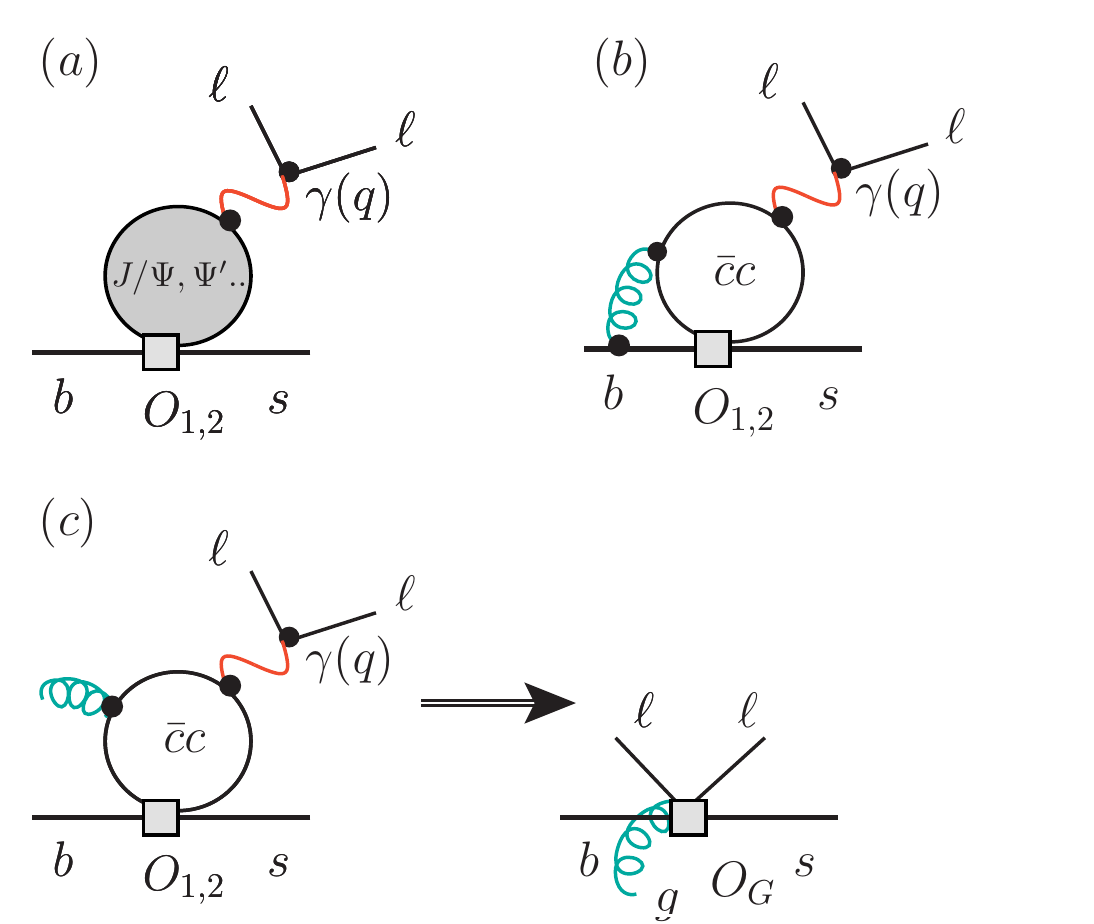}
\caption{\small  Contributions to high-$q^2$ OPE. (a) dimension $3$ factorisable correction (b) dimension $3$ example of ${\cal O}(\al_s)$-correction (c) dimension $5$ operator shown with loop and contracted loop.  }
 \label{fig:OPEc}
 \end{center}
 \end{figure}

The correlation function that contributes to the helicity amplitude $\HH^V$ \eqref{eq:hV}  is given by
\begin{equation}
\Gamma_\mu =  \sum_k C_k \int d^4 x e^{i q \cdot x} \matel{K(p)}{T j_\mu^{\rm em}(x) \Op_k(0) }{B(p_B)} = Q_c \Gamma^{(c)}_\mu +  Q_s  \Gamma^{(s)}_\mu + Q_b  \Gamma^{(b)}_\mu   \;,
\end{equation}
where $\Op_k$ is one of the four quark operators in \eqref{eq:SMbasis} and $ j_\mu^{\rm em}$  is the electromagnetic operator.
In this work we are only interested in effects  that contribute to a single 
resonant structure in the $\bar cc$-channel. Hence we restrict our attention 
to the electromagnetic charm current contribution $\Gamma^{(c)}_\mu$. 
Three typical contributions to  $\Gamma^{(c)}_\mu$  are indicated in 
 Fig.~\ref{fig:OPEc}. Fig.~\ref{fig:OPEc}a  corresponds to the FA studied in section \ref{sec:fac}. 
The contributions in Fig.~\ref{fig:OPEc}bc  are the kind of contributions whose size we  intend to assess  in this section.  

To do so we extend $C_{7,9}^{\rm eff}$  to include the correction from 
\cite{GPS08}
\begin{equation}
C_9^{\rm eff} =  C_9 + \delta C_9^{\rm \fac} + \delta C_9^{\rm \cor}   ,  \quad  C_7^{\rm eff} = C_7 +  \de C_7^{\rm \cor}  \;,
\end{equation}
where $\delta C_9^{\rm \fac} =  \cnf h_c(q^2) $ (Fig.~\ref{fig:OPEc}a) was implicitly given in \eqref{eq:hV}  and $\delta C_{7}^{\rm \fac}  = 0$.  The non-factorisable contributions (Figs.\ref{fig:OPEc}bc) are denoted by $\de C_{7,9}^{\rm \cor}$. The correction 
to Fig.~\ref{fig:OPEc}b is given by 
$\delta C_{7,9}^{\rm \cor}|_{\rm Fig.\ref{fig:OPEc}b} = 
- \frac{\al_s}{4\pi} (C_1  F_1^{(7,9)}(q^2) +C_2 F_2^{(7,9)}(q^2))- \al_s \cnf h_c^{(1)}$. 
We have subtracted the  $\al_s$-corrections $h^{(1)}_c$  from the result $F_i^j$ given in \cite{GPS08}.\footnote{Note \cite{GPS08} uses the basis \cite{Chetyrkin:1996vx} which differs from the one used throughout this paper}  The contribution $h^{(1)}$  corresponds to fig1e in \cite{GPS08}.
In order to compare the relative size  we introduce  the ratio
\begin{eqnarray}
\label{eq:xs}
x(s)  =  \frac{\HH^{V,\cor}(s)} {\HH^{V,\fac}(s)} =
\frac{\delta C_9^{\cor}(s) + \delta C_7^{\cor}(s)\varphi(s)}{ \cnf h_c(s)}  \;.
\end{eqnarray}
%|_{\rm Fig.\ref{fig:OPEc}b(c)}}
The function $\varphi(s) = (2m_b f_T(s) )/((m_B +m_K )f_+(s))$ is the quotient  of the $C_{7,9}^{\rm eff}$-prefactors in \eqref{eq:hV} and is close to unity and slowly varying. 
Ideally we would like to know the function $x(q^2)$ through the high $q^2$-region in which case we would simply incorporate it into the results. As emphasised above we cannot hope
to do that since $\HH^{V,\cor}(s)$ only makes sense when integrated over a suitable (duality) interval.

\begin{itemize}
\item  Diagram Fig~5B: vertex corrections \\ 
These  have been evaluated for the inclusive decay $b \to s \ell \ell$, in the high $q^2$-region, numerically  \cite{GIHY03} and analytically \cite{GPS08}, in a well-converging expansion in $m_c^2/m_b^2$.  
Using the work of \cite{GPS08,merci2} we were able to extract the $\Gamma^{(c)}$-contribution. 
The subtraction of $h^{(1)}$ as described above  leads to a sizeable enhancement 
of $H^{V,\cor,b}$ at $\mu = m_b$. It is found that  $|x(s)| \simeq 0.5$. Hence this contribution is sizeable. 
 In section \ref{sec:partonic} this estimate will be tested against cancellation effects under the duality integral.  

\item  Diagram Fig~5C: $\bar b G s$-corrections \\
 The  correction in Fig.~\ref{fig:OPEc}c  results in a dimension five matrix element  of the form
$\matel{K}{\bar s_L \gamma_\la g G_{\al \be}b}{B}/q^2 f(4m_c^2/q^2)$ with appropriate contractions of kinematical factors. The function $f$ originates from the charm-loop and has a form similar to  
$h^{(0)}_c(q^2)$  \eqref{eq:h0} in leading order perturbation theory.

The remaining gluon can either connect to  the spectator $\HH^{V,\rm spec}$,  the $K$-meson $\HH^{V,gK}$  or 
 the $B$-meson  $\HH^{V,gB}$. The first case has been evaluated in \cite{BBF11}  within QCD factorisation.  Comparing the latter  with $\HH^{V,\fac}(q^2 > \qref)$  translates 
into  $|x|_{Fig.~\ref{fig:OPEc}c}^{\rm spec} \simeq 0.02$. 
%Incidentally  the relation ${\rm Disc}[\HH^{V,\cor,c}](s) = 2i \im[ \HH^{V,\cor,c}](s)$ holds as there are not other cuts than the one in $q^2 > 4m_c^2$. 
Even though QCD factorisation can only give a rough estimate in the region, as emphasised by the authors of \cite{BBF11}, this strongly indicates that this type of correction is negligible; especially as compared to the vertex corrections.  Note this is also consistent with the hard spectator scattering contributions in \cite{Beneke:2001at} being roughly $4\%$ (at $q^2 = 8 \GeV^2$) as compared to the leading order contributions.
%Other corrections come from the gluon being connected with either the $K$-meson 
%or the $B$-meson which we shall denote by  $\HH^{V,gK(gB)}$ respectively. 
We have evaluated $\HH^{V,gK}$ in light-cone sum rules at $q^2 \simeq \qref$ and find 
\cite{LZ14b} $|x|_{Fig.~\ref{fig:OPEc}c}^{\rm gK} \simeq 0.02$.\footnote{The light-cone expansion for the Kaon might give a reasonable value for $q^2 \simeq 15 \GeV^2$.}
\footnote{This effect is of importance since, besides $m_s$-corrections, as it constitutes the leading correction to the helicity hierarchy in $B \to V \ell \ell$-decays, relevant to the search of right-handed currents.}
The contribution $\HH^{V,gB}$ has been assessed in \cite{KMPW10} in the low $q^2$-region and comparing with  \cite{MXZ08} it is found that $  |\HH^{V,gB}(0)| \simeq 2 |\HH^{V,gK^*}(0)|$ which indicates that this contribution is negligible as well. 
We wish to emphasise that it might be worthwhile 
to check the size of the different contributions in one framework at $O(\al_s)$ rather than gathering results from three different approaches.

\end{itemize}

In summary we have identified the vertex corrections Fig.~\ref{fig:OPEc}b as the main source of corrections.
Since we are really interested in the local $q^2$-behaviour of the corrections we need to go further and address 
the question in the hadron picture through quark hadron duality.

\subsection{Dispersion relations, (non)-positivity and quark hadron duality} 
\label{sec:dispersion}

The canonical approach to quark hadron duality is based on  dispersion relations e.g. \cite{PQW,Shifman} which follow from 
first principles.
Dispersion relations are well established at the amplitude level and in essence just require knowledge 
of the analytic structure on the physical sheet. 
%Hence we have to content ourselves with an error estimate of this contribution.
%For our purposes it is sufficient to know whether $\HH^{V,\cor}(q^2)$ can turn the qualitative effect of the resonances around.
%In order to assess this question we propose  to reason in a slightly different way. 
We assume\footnote{To be justified in subsection \ref{sec:partonic}.} that the helicity amplitude $\HH^{V,\cor}(q^2)$ has the same analytic structure as 
$\HH^{V,\fac}(q^2)$ ($h_c$ respectively) and therefore obeys the same dispersion relation \eqref{eq:dispersion}
\begin{equation}
\label{eq:dis2}
\HH^{V,X}(s) = \HH^{V,X}(s_0)  + \frac{(s-s_0)}{2 \pi i} \int_{\cut}^{\infty} \frac{dt}{t-s_0} \frac{{\rm Disc}[\HH^{V,X}](t)}{t-s-i0} \;,  \quad X \in \{ \fac,\cor \} \;.
\end{equation}
 %In this work we are interested in the local behaviour of the charm-resonances. 
%Somewhat more modestly we want to assess whether the LHCb-spectrum is in any way compatible with the partonic vertex corrections. 
The local behaviour of $\HH^{V,X}(q^2)$  near a resonance $r$  is well  approximated by a Breit-Wigner resonance\footnote{The Breit-Wigner form is a good approximation near the resonance in a range  governed by the width. The Breit-Wigner ansatz cannot be a good approximation everywhere since 
it has got a pole on the physical sheet at $ q^2= m_r^2 -i m_r \Gamma_r$ in contradiction with 
the analytic structure of the K\"all\'{e}n-Lehmann representation. 
This deficiency though does not matter as long as one stays in the range mentioned above.}   
\begin{equation}
\label{eq:BW}
\HH^{V,X}(q^2 \simeq m_r^2) \simeq \frac{- r^X_r}{q^2 - m_r^2 + i m_r \Gamma_r} + .. \;,
\end{equation}
where we shall refer to $r^X_r$ as the residue.
A first important point is that\footnote{\label{foot:amp} More precisely $r_r^{\fac} + r_r^{\cor} \propto
{\cal A}(B \to K r)|_{O_{1,2}}  {\cal A}(r \to \ell \ell)$, where ${\cal A}$ stands for the amplitude and $O_{1,2}$ indicates a restriction of the effective Hamiltionian to these operators.}
\begin{eqnarray}
\label{eq:r}
r_r^{\fac} &=&  \cnf \frac{3 \pi}{\al^2} \Gamma^{r \to \ell\ell} m_r > 0  \;,   \quad  \nonumber \\
r_r^{\cor} &\in& \mathbb{C}  \;.
\end{eqnarray}
For $r_r^\fac $ we have quoted the result of the Breit-Wigner approximation.  
 Positivity of $\Disc[H^{V,\fac}]/(2 i)= \im[H^{V,\fac}] $  follows on more general grounds. 
 First from the positivity of the cross section $ R_c(s)>0$.  Second from the   positivity of the  K\"all\'{e}n-Lehmann  spectral representation 
of a \emph{diagonal}  two point  function.
 For  $r_r^{\cor}$ there is no such constraint and $r_r^{\cor}$ is generally a complex number (as in fit-d)  where the phase is associated with the scattering phase of the 
corresponding amplitudes (c.f. footnote \ref{foot:amp}).  Hence the major pitfall we have to be concerned with is that due 
to non-positivity of the $r_r^{\cor}$ (or $\Disc[H^{V,\cor}]/(2 \pi i))$  a global dispersion integral might majorly underestimate local or semi-local effects.  
The fact that all the residues in fit-c) come out with the same sign  suggests that this is presumably not the case. On the other hand fit-d) indicates that there could be 
cancellations to some degree as the phase varies (mildly) as a function of $q^2$.
For this reason we have to further pursue our investigation with some care and detail.

\subsubsection{Intervals of duality in $\BKll$}

\begin{figure}[h!]
 \includegraphics[scale=1.00]{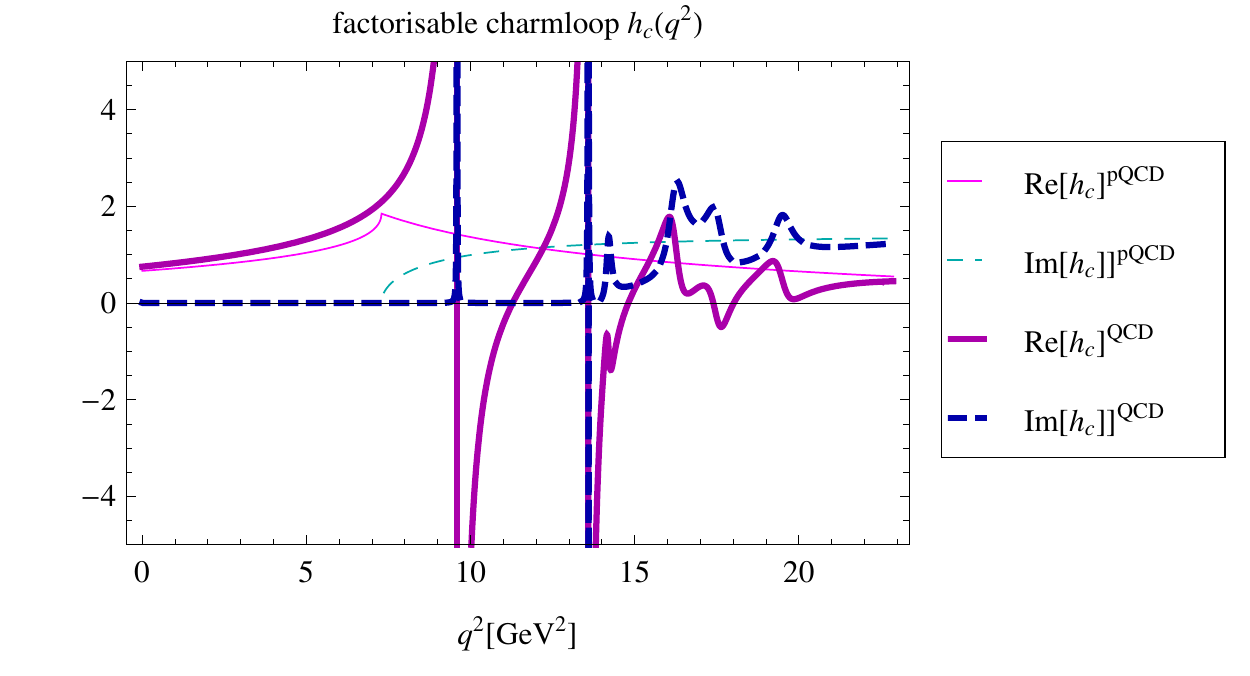} 
 \caption{\small Plots of the real (straight purple line) and imaginary part (dashed blue line) of the perturbative QCD 
 $h^{(0)}$-function (thin lines) and the QCD $h_c$ (thick lines) as fitted from BES data. The latter are shown in Fig.~\ref{fig:h}. $\re[h_c^{(0)}]$ is obtained through the dispersion integral \eqref{eq:dispersion}. 
 %The purpose of this figure is aim the illustrate the discussion in the text.
 }
 \label{fig:deux}
 \end{figure}

\begin{figure}[h!]
 \includegraphics[scale=0.60]{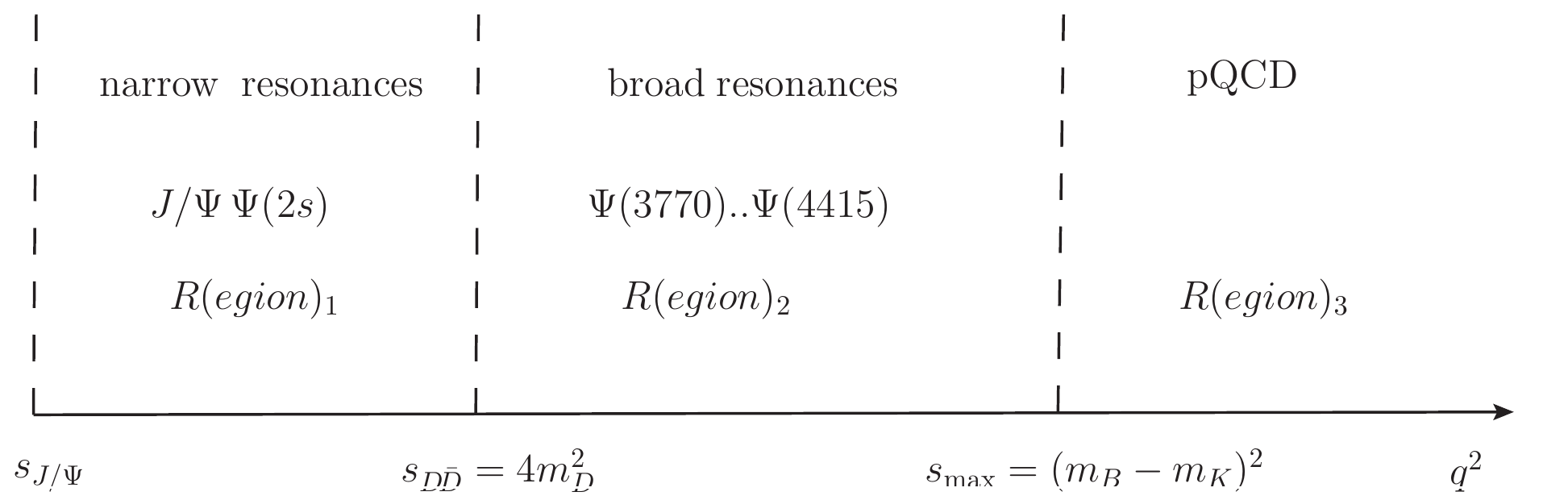} 
 \caption{\small Three regions in $q^2$, relevant to the charm-resonances: narrow resonance region which cannot be described by perturbative QCD (left), broad resonance region, of interest to this work, described by perturbative QCD on average (middle) and the third region which is even locally described by perturbative QCD (right).}
 \label{fig:region}
 \end{figure}

Using that below the thresholds 
\begin{equation}
\label{eq:global}
\HH^{V,X}(s)^\hadron \simeq   \HH^{V,X}(s)^\parton  \;, \quad s \ll 4 m_c^2 \;,
\end{equation}
fixes the problem of the subtraction point \eqref{eq:dis2}.
Global quark hadron duality in (\ref{eq:global},\ref{eq:dispersion}) translates into 
\begin{equation}
\label{eq:d1}
\aver {\Disc[\HH^{V,X}]^\hadron}_{\bar \omega}^{(\cut,\infty)} \simeq \aver {\Disc[\HH^{V,X}]^\parton}_{\bar \omega}^{(4 m_c^2,\infty)}
\end{equation}
 provided 
  the weighting function 
\begin{equation}
\label{eq:omegaD}
\bar \omega(t) \equiv \frac{1}{(t-s_0)(t-s-i0)}   
\end{equation} 
is 
used under the integral average
\begin{equation}
\label{eq:average}
\aver{f}^{(s_1,s_2)}_{\omega} \equiv \int_{s_1}^{ s_2} dt \omega(t) f(t) \;.
\end{equation}
Relation Eq.~\eqref{eq:d1} is precise up to the order  $\al_s$ to which the right hand side is computed 
in perturbation theory.
It is self understood that the variables $s$ and $s_0$ are sufficiently far away from the discontinuities. 
The crucial question, known as \emph{semi-global quark hadron duality} e.g. \cite{Shifman}, is then to what degree this equation still holds when the integration interval is split up into different regions. For our purposes the discussion naturally splits into three regions shown in Fig.~\ref{fig:region}:   
the lowest interval from  $s_{J/\Psi}$ to   the $D\bar D$-threshold $\qref \equiv 4 m_D^2 \simeq (3.73 \GeV)^2$ (including the two narrow resonances $J/\Psi$ and $\Psi(2S)$),
the interval therefrom to the kinematic endpoint $ s_{\rm max} \equiv (m_B-m_K)^2  \simeq (4.75 \GeV)^2$ and the third interval extending to infinity. We shall denote those three regions by $R_{1,2,3}$ respectively.  In the interval $R_3$, not accessible in $B \to K$, the resonances 
are very broad and smoothly become a part of the continuum.  This region is well described by perturbative QCD even locally.  
This can be inferred from the comparison with the BESII-data for $\Disc H^{V,\fac}$ \cite{KST07}.
 It therefore follows that
\begin{equation}
\label{eq:d2}
\aver {\Disc[\HH^{V,X}]^\hadron}_{\bar \omega}^{(\cut, s_{\rm max}  )} \simeq \aver {\Disc[\HH^{V,X}]^\parton}_{\bar \omega}^{(4 m_c^2, s_{\rm max})} \;,
\end{equation}  
is a reasonable approximation (semi-global quark hadron duality).
 Whereas the local description of the narrow resonance region $R_1$ is particularly hopeless, the same 
is not true of the region $R_2$ where the $D\bar D$-threshold renders the resonances sufficiently broad such that 
perturbation theory becomes valid on average. This is at least true for $\Disc H^{V,\fac} = \cnf h_c$, as can be inferred from the plots of Fig.~\ref{fig:deux}  or the equivalent plots of the $R$-function 
in \cite{PDG}.  Hence one ought to  expect that 
\begin{equation}
\label{eq:d3}
\aver {\Disc[\HH^{V,X}]^\hadron}_{\bar \omega}^{(\qref , s_{\rm max}  )} \simeq \aver {\Disc[\HH^{V,X}]^\parton}_{\bar \omega}^{(\qref, s_{\rm max})} 
\end{equation} 
holds approximately. Eqs.~ (\ref{eq:d1},\ref{eq:d2},\ref{eq:d3}) are expected to hold for  smooth smearing function $\omega$, other than $\bar \omega$ \eqref{eq:omegaD} \cite{PQW}.  For example for the  penguin amplitude it interferes with in $\BKll$. 
Relation \eqref{eq:d3} is the basis of further investigations. 
We shall use \eqref{eq:d3}  to put the previous quoted  estimate $|x| \simeq 0.5$ on more solid grounds. In a second step  apply it directly to the extracted fits to the LHCb-data.
 
\subsection{The size of the SM vertex corrections  over duality interval}
\label{sec:partonic}

The residue version of factorisable versus non-factorisable contribution \eqref{eq:xs} is given by
\begin{equation}
x_r = (\rho_r -1) = \frac{r_r^{\cor}}{r_r^{\fac}} \simeq  \left. x_D(s) \right|_{s \simeq m_r^2} \;.
\end{equation}
where
\begin{equation}
\label{eq:xDs}
x_D(s) \equiv \frac{ {\rm Disc}[\HH^{V,\cor}](s)}{{\rm Disc}[\HH^{V,\fac}](s)} \;.
\end{equation}
The quantity $x_D$ is an improved version of $x$ \eqref{eq:xs} since it is does not depend on 
the subtraction constant for example which is immaterial to the shape in the region of interest.
The contribution of the FA  is given by  ${\rm Disc}[\HH^{V,\fac}](s) = 2i \im[ \HH^{V,\fac}](s)$ \eqref{eq:3},
whereas the function $ {\rm Disc}[\HH^{V,\cor,b}](s)$ does not obey such a simple relation since there are cuts below the charm threshold
 $\im [\HH^{V,\cor,b}]_{q^2 < 4 m_c^2} \neq 0 $. For example cuts  
 in the variable  $m_b^2 \geq (2m_c+m_s)^2$ in Fig.~1c in \cite{GPS08}. 
Using the results in \cite{AACW01} 
one can verify that $\im[H^{V,\cor,b}](4m_c^2 - \Delta)/ \im[H^{V,\cor,b}](4m_c^2 + \Delta)$ (for $0 < \Delta < 2\GeV^2$) is a very small quantity and hence those cuts are negligible in the region where 
$q^2 > 4m_c^2$.\footnote{This assertion is true beyond the finite gap due to the Coulomb singularity originating from 
diagram 1e in \cite{AACW01,GPS08}. In any case this diagram corresponds to the $h^{(1)}$-correction in \eqref{eq:hPT} which we do subtract as explained previously.}  
Hence we conclude that 
\begin{equation}
\label{eq:super}
{\rm Disc}[\HH^{V,\cor,b}](s) \simeq 2i \im[ \HH^{V,\cor,b}](s)
\end{equation}
 is a good approximation.\footnote{We have verified this chain of arguments by using 
\eqref{eq:super} in a dispersion relation of the type \eqref{eq:dispersion}. 
There is a further complication. The results in \cite{GPS08} are not valid for $q^2 > m_b^2$.  
We have overcome this problem   by setting the function to its value at $q^2 = m_b^2$ for $q^2 > m_b^2$.  This  ought to be a good approximation since the function is expected to go to a constant corresponding  to the logarithmic UV-divergence. The result 
agrees extremely well for $s$ close to the subtraction point $s_0$ which justifies our previous assertions. 
A further benefit is that the explicit (approximate) construction of the dispersion relation also eliminates doubts about complex anomalous thresholds which can appear on the physical sheet 
in $\BKll$-type decays e.g. \cite{O8}. Complex anomalous thresholds would invalidate Eq.~\eqref{eq:super}.} Once more we emphasise that only the vertex corrections proportional to $Q_c$ are to be considered. The procedure for 
obtaining them has been outlined under the first item in section \ref{sec:charm_out}.
The quantity $x_{D}^b \simeq - 0.5$ 
throughout the relevant interval $\qref < q^2 <  \qmax$ for $\mu = 4 \GeV$ and only slightly higher values for 
$\mu= 2 \GeV$ as can be inferred from Fig.~\ref{fig:scale} in appendix \ref{app:scale}.  
The optimal choice of scale $\mu = m_b$, aimed at maximising the effect of the BESII-data, 
is discussed in appendix  \ref{app:scale}.
 Hence the size of the vertex correction integrated over $R_2$ (c.f. Fig.~\ref{fig:region}) is 
 approximately given by
 \begin{equation}
 \label{eq:xDSM}
 (x_D)^{\rm SM}_{R_2} = \frac{\aver { {\rm Disc}[\HH^{V,\cor}]^\hadron}^{(\qref,s_{\rm max}) }} {\aver { {\rm Disc}[\HH^{V,\fac}]^\hadron}^{(\qref,s_{\rm max}) }  }  
 \stackrel{\eqref{eq:d3}}{\simeq} 
 \frac{\aver { {\rm Disc}[\HH^{V,\cor}]^\parton}^{(\qref,s_{\rm max}) }} {\aver { {\rm Disc}[\HH^{V,\fac}]^\parton}^{(\qref,s_{\rm max}) }  }   \simeq -0.5  \;,
 \end{equation}
 for any reasonably smooth smearing function $\omega$.
The last equality follows from the fact that $x_D(s)$ is nearly constant throughout the region $R_2$. 
This is estimate puts the previous estimate $|x| \simeq 0.5$ on more solid grounds and settles the effect 
of the sign in a more definite way. 
As previously mentioned a  correction of $-0.5$ \eqref{eq:xDSM} is down by a factor of seven.  
%with respect to the shift $-3.5$ suggested by fit-b) ($\n = -2.55$).  
%In the next section we assess whether the strong phases can change this conclusion considerably.

\subsection{The local and semi-global charmonium excess over FA from LHCb-data}
\label{sec:data-fit}

In this subsection we perform a similar analysis as before but comparing the actual discontinuity of the amplitudes that we 
extract from the fit. We would like to test whether (non)-positivity can lead to effects that are underestimated in duality integrals. 
An important aspect is that the fits c) and d) do not contain a background model, a problem we have commented on in section \ref{sec:combined}, and hence we can only extract the resonant contribution. 
The background is expected to be smooth and does therefore not influence the resonant shape in any 
significant way.

\begin{figure}[h!]
 \includegraphics[scale=0.9]{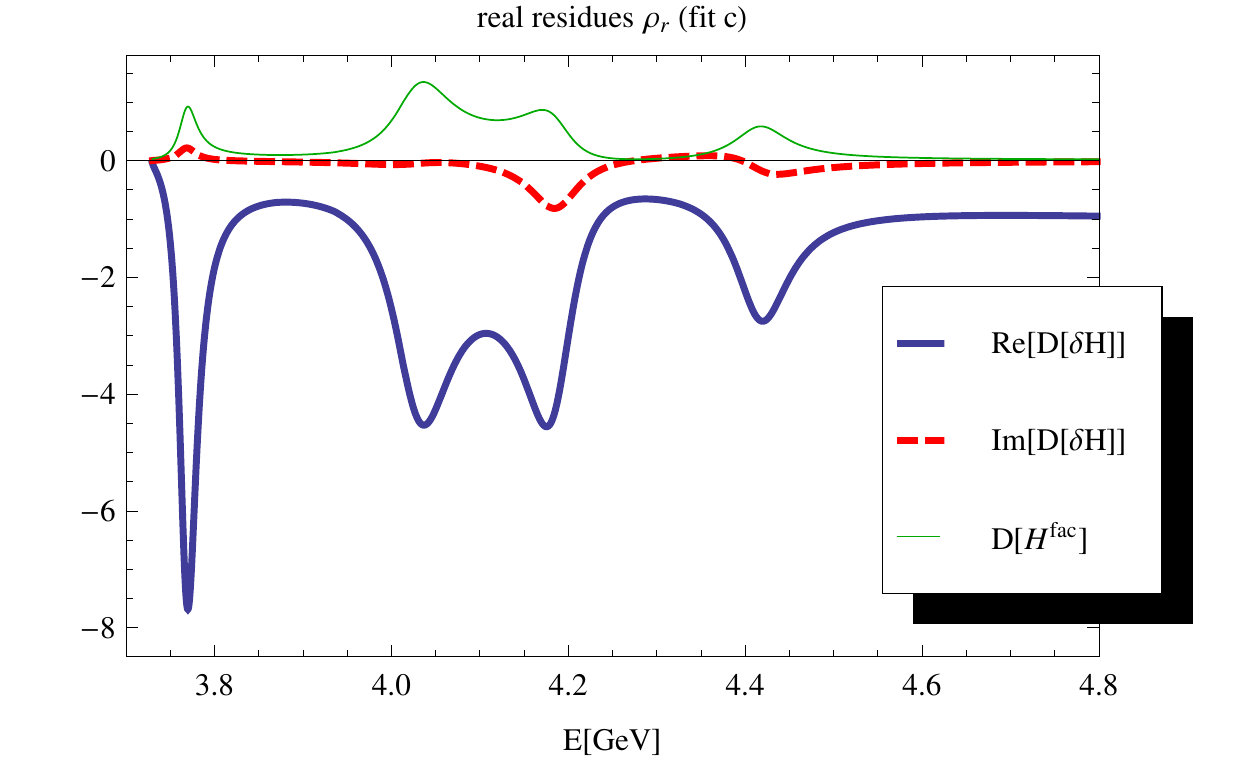} \\
   \includegraphics[scale=0.9]{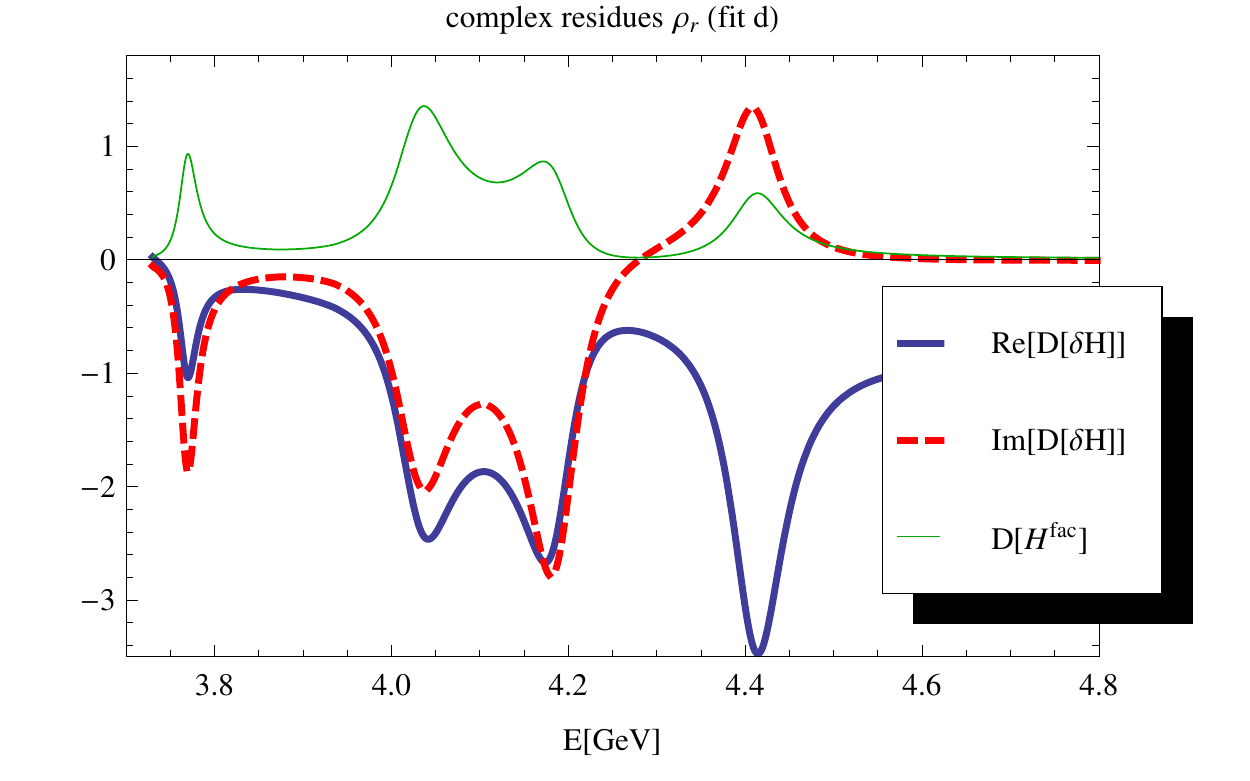} 
 \caption{\small Plots of fits c) (top) and d) (bottom) with real and complex residues.  
The abbreviation in the plots stand for the following shorthands  $    \re[D[\delta H]] + i \im[D[\delta H]] =  2i \cnf \Disc [ \delta H_{\rm res}^{V,\rm fit} ]   $ \eqref{eq:Dd} (thick blue and red dashed line for imaginary and real part)
and   $  = D[H_{\rm res}^\fac] = 2i \cnf \Disc [ H_{\rm res}^{V,\fac} ]  $ (thin green line) with $\cnf(m_b) \simeq 0.6$. The plots are a direct reflection of the fit-data given in table \ref{tab:fit1}.}
  \label{fig:realcomp}
 \end{figure}

Hence the discontinuity of the resonant part beyond FA, \emph{as extracted from the fit}, is given by
\begin{equation}
\label{eq:Dd}
\Disc [ \delta H_{\rm res}^{V,\rm fit} ]  = \Disc [ H_{\rm res}^{V,\rm fit}]    - \Disc [H_{\rm res}^{V,\fac} ]  \;,
\end{equation}
where  
\begin{eqnarray}
\label{eq:Dfac}
& & \Disc [H_{\rm res}^{V,\fac}] = \cnf \frac{6 \pi i}{ \al^2} \sum_f |\sum_r T^{r \to f}|^2  \; , \\[0.1cm]
& &  \Disc  [ H_{\rm res}^{V,\rm fit} ]  =   \cnf \frac{6 \pi i}{ \al^2}   \sum_f ( \sum_r  \rho_r T^{r \to f }(s)   )  ( \sum_r T^{r \to f }(s)   )^*   \;,
\end{eqnarray}
follow from Eqs.~(\ref{eq:R},\ref{eq:3}) and Eq.~\eqref{eq:rhor} respectively. 
Note that $ \Disc [ H_{\rm res}^{V,\rm fit}]$ is generally not real, even for real $\rho_r$. Plots 
of the various quantities, which can be reconstructed from the fit-data in  table \ref{tab:fit1}, are given in Fig.~\ref{fig:realcomp}. 
For the fit-c) there are no significant signs of cancellation effects whereas for the fit-d)  we can see that the imaginary part does cancel to some extent when integrated over the interval as an effect of the
 approximately $90^\circ$ phase shift between the residues $\rho_{\Rc}$ and $\rho_{\Rd}$. 
This shift might not be a solid feature since  the model, in that region, does not result in a very good   fit  of the LHCb rate (c.f. Fig.~\ref{fig:rate2}). 
Finally we perform the averages of the duality interval to find 
\begin{equation}
\label{eq:xDres}
 (x_D)^{\rm res-data}_{R_2} =  \frac{\aver { {\rm Disc}[\delta \HH^{V,\rm fit}_{\rm res}]}^{(s_{\rm th},s_{\rm max}) }_{\omega=1}} {\aver { {\rm Disc}[\HH_{\rm res}^{V,\fac}]}^{(s_{\rm th},s_{\rm max}) }_{\omega=1}  }     \simeq   
  \begin{cases}  4.0 e^{-i 176^{\circ}}   \simeq 3.2 e^{- i176^{\circ}}  [\nn{c}^{-1}]    &  \text{fit-c)}  \\[0.1cm]
                          2.5 e^{-i 152^{\circ}}   \simeq 2.7 e^{- i152^{\circ}}  [\nn{d}^{-1}]   &   \text{fit-d)}      \end{cases}  \;.
  \end{equation}
The numbers are robust under change of smooth smearing function, for example for $\omega(q^2) = 1/(q^2(q^2+ 8 \GeV^2))$ we get $4.1 e^{- i177^{\circ}}$ and  $  2.5 e^{-i 147^{\circ}}$ for fit-c) and -d) respectively.
The global scaling factor $\n$ \eqref{eq:fp1} refers to the number in table \ref{tab:fit1} which has to  
be taken into account when comparing numbers between different fits.
We observe a shift 
from $3.2$ to $2.5$ due to complex residues which is what one would expect by inspecting the graphs in Fig.~\ref{fig:realcomp}. 
This corresponds to an effect below $25\%$ and can not be seen as very significant. The numbers $3.2 [\nn{c}^{-1}] $ and 
$2.5 [\nn{d}^{-1}]$ have to be compared with $|\cc-1| \simeq 3.5[\nn{b}^{-1}] $ and are a bit but not really significantly lower. 
One has to keep in mind that, for fits-c) and -d), no background model has been used to fit the LHCb-data 
as previously explained.  
In this sense the relative closeness of the results of fit-c) and fit-d) are more important than the relative closeness to  fit-b).

\subsection{Summary of assessment of non-factorisable corrections}
\label{sec:summary}

In assessing the non-factorisable contributions in the SM model we have made use 
of the large scale $q^2 \geq \qref$ by integrating out the charm quarks. 
The vertex corrections in Fig.~\ref{fig:OPEc}c were identified as sizeable. 
The relative size of corrections were found to be $-0.5$ with respect to the factorisable corrections  when integrated  over the duality region $R_2 = [\qref,\qmax]$.
This is substantially below  -$3.5$ as suggested by fit-b) in table \ref{tab:fit1}. 
The conclusion remained robust under fits c) and d) when the residues were allowed to depart from 
the ratios dictated by the FA.  The data does not suggest major cancellations, due to non-positivity of the duality integrand, in the case of complex residues c.f. \eqref{eq:xDres}.  
In our assessment we have not found any signs of sources within  the SM that could give rise to such large corrections.

\section{Consequences, strategies and  speculations on the origin of the charmonium anomalies}
\label{sec:strategyandspec}

In the previous section we have analysed whether or not the $\BKll$ charm-resonances 
can be accommodated within the dimension six effective Hamiltonian \eqref{eq:effective-ew-hamiltonian}, commonly used 
to describe $b \to s \ell \ell$-transitions. We have found no indications that this is the case with current understanding.  In subsection \ref{sec:strategy} we propose strategies to measure the effect in other 
$b \to s\ell\ell$ observables and or transitions. 
One of the main goals  being to extract the opposite parity Wilson coefficient combination.
In  subsection \ref{sec:lowq2} we investigate the connection to the $B \to K^* \ell \ell$-anomalies of the year 2013. This is not an obvious task since we do not have any precise knowledge of the microscopic effect that leads 
to the anomalous resonance behaviour.
Finally in subsection \ref{sec:newOP} we briefly  entertain speculations beyond the SM. 
The essence of the discussion is summarised in subsection \ref{sec:summary-end}.

\subsection{Strategies to disentangle the microscopic origin of the charm-resonance anomalies}
\label{sec:strategy}

First we note that, on grounds of parity, the $\BKll$-transition couples to the vector current and therefore 
to the $ C_+ \equiv C + C'$-Wilson coefficient combination. Assuming that the $\ell\ell$-pair emerges through a photon  the effect can be absorbed into the $C^{\rm eff}_{9+} \equiv C_9^{\rm eff} +  C_9^{\rm eff}$ Wilson coefficient combination. 

\begin{figure}[h!]
{ \includegraphics[scale=0.64]{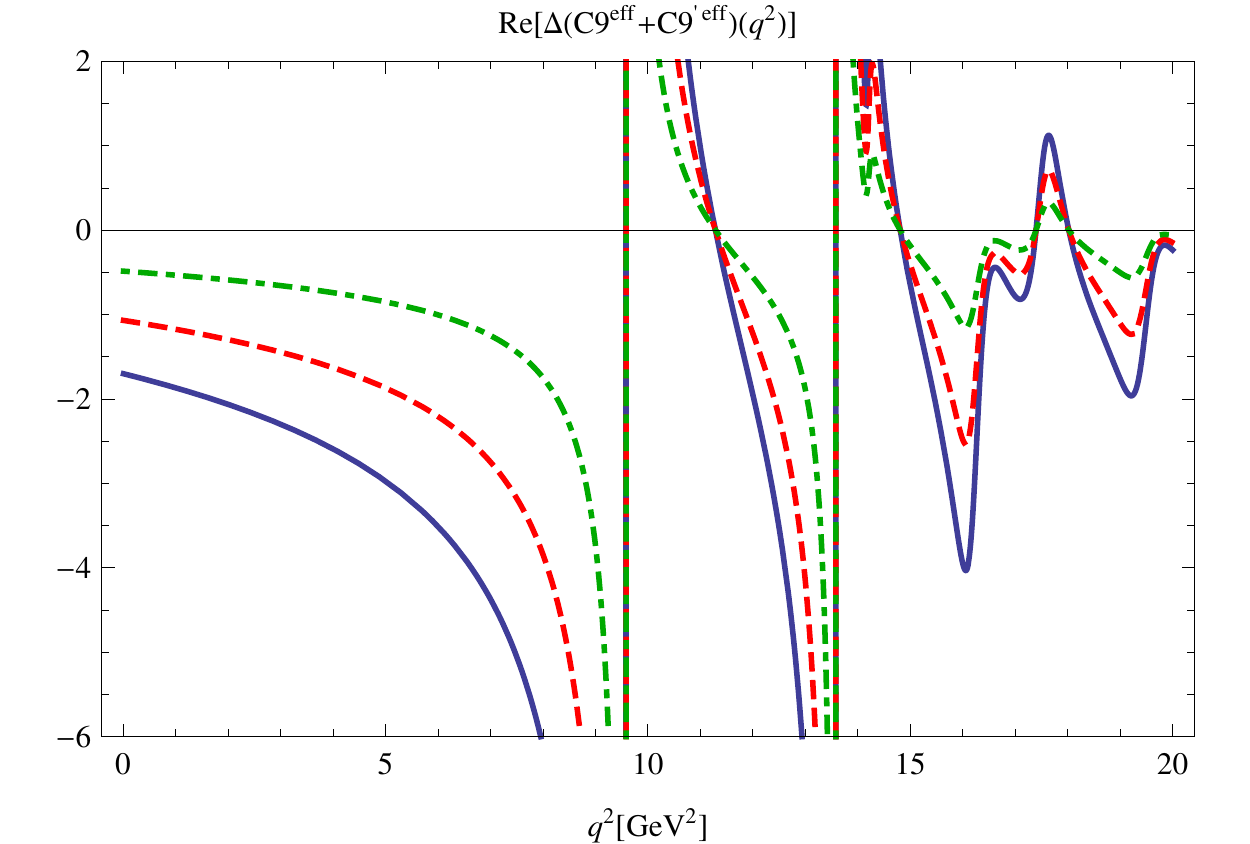} \;
 \includegraphics[scale=0.7]{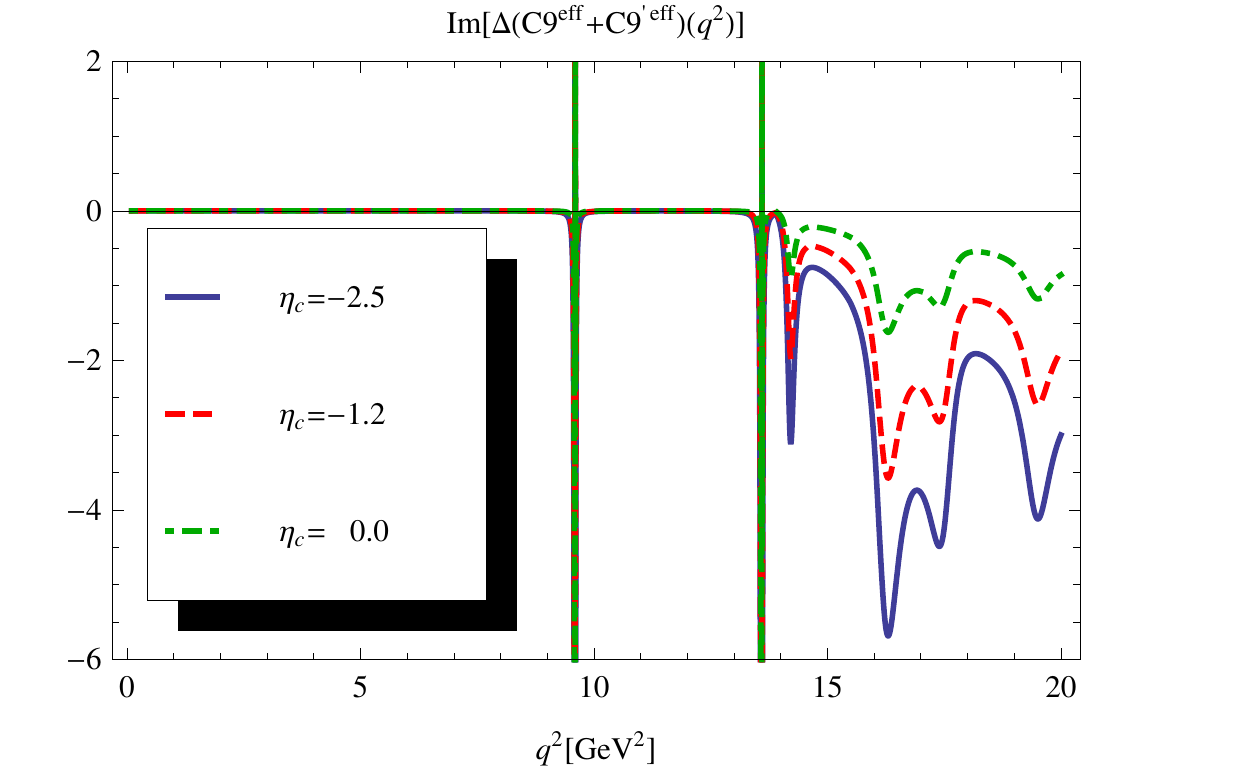} }
 \caption{\small Real and imaginary part of $\Delta C_9^{\rm eff}(q^2)$  
 ($C_9^{\rm eff}(q^2) \equiv C_9^{\rm eff,SM}(q^2) +  \Delta C_9^{\rm eff}(q^2)$)  for different values of $\n = -2.5,-1.2,0$. 
 The value $\cc  = -2.5$ corresponds to the fit-b) in section \ref{sec:combined}.  }
 \label{fig:C9eff}
 \end{figure}

In order to assess the nature of the effect we are going to use  fit-b) in table \ref{tab:fit1} as a template and simply scale the factorisable part by  factors of $\cc$. We shall refer to this type as the $\cc$ scaled-FA.
The effect on $C^{\rm eff}_{9+}$ is shown for real and imaginary part  in Fig.~\ref{fig:C9eff}. 
There are several reasons why we choose fit-b) over fits c) and  d). 
First for fit-b) we, at least, have a microscopic (effective) theory at hand whereas for the other two fits this is not the case.  In addition, and of course related, for fits c) and  d) we did not incorporate  a background model and a subtraction constant.  As previously mentioned the drop in $\chi^2$ is not overwhelming c.f. Tab.~\ref{tab:fit1} and this underlines the importance of the first remark.
One has to keep in mind that  only one observable, namely the high-$q^2$ $\BKll$-rate, was fitted for.
 With future data the situation would improve considerably.\footnote{ In particular the release of the $J/\Psi$- and $\Psi(2S)$-data could be of use in order to assess the strong phase of the amplitude. 
 The magnitude can be obtained from $B \to J/\Psi K$ and $B \to \Psi(2S) K$.}  
 First one could hope  to fit for a background model and a (real) subtraction constant in \eqref{eq:d2}. 
 Second   the plethora of observables in $b \to s \ell \ell$  constrain the ${\rm Disc}[\HH^{V,X}]$ 
 in \eqref{eq:d2} more severely.  In essence the plots in this section should be seen as an illustration of 
 the effect and we therefore, al least in this version,  do not give uncertainties for the predictions. 
 Yet the $\cc = (1,0)$ case corresponds to the SM-FA and can be discriminated against future experimental data. 
%The most important question, of whether or not, the SM or QCD respectively, can be held responsible for the observed phenomenon, could be answered to the negative if one establishes a sizeable correction in $C_9^{'\rm eff}$.
%This kind of effect can be accommodated in $C_9^{\rm eff}(q^2)$. 
We list a few remarks relevant to $\BKll$ and $\BKsll$ (and or $B_s \to \phi \ell \ell$\footnote{It would seem that for all practical purposes $B_s \to \phi \ell \ell$ is on an equal footing to $\BKsll$ since the spectator quark is not expected to play a relevant part in all of this.}). First $\BKll$:
\begin{itemize}
\item $C_{9+}$  decreases the $\BKll$ branching fraction in the low and 
high $q^2$-region which is in qualitative accordance with the recent LHCb analysis with larger binning  
\threefb \cite{LHCb14Iso}.  For high $q^2$ this is of course only consistent with the finer binning result 
elaborated on in this work. For low $q^2$ the decrease arises through the decrease of 
$C_{9+}$.
\item The shift of  $C_{9+}$  on average is of the same order as demanded 
by the $P_5'$-anomaly \cite{understanding,AS13,latticepheno,Beaujean:2013soa}. 
There is though an important qualitative difference in that the shift, suggested by our work, is 
$q^2$-dependent rather than a uniform shift  \cite{understanding,AS13,latticepheno,Beaujean:2013soa}. 
More comments can be found in subsection \ref{sec:lowq2}.
\item The two other angular observables in $\BKll$ (due to the opening angle of the lepton-pair), $F_H$ and $A_{\rm FB}$ 
are proportional to effects of ${\cal O}(m_l)$ in the SM; e.g. \cite{BHP07}.  
It therefore seems, currently, difficult 
to extract sensible information in our framework. The LHCb-data at  
\threefb \cite{LHCbBKllang} is consistent with 
the tiny SM predictions. The two observables are of course of  importance  
to set bounds on new physics operators.
%\begin{equation}
%\frac{d \Gamma}{\Gamma d \theta_{\ell}} =   F_H + 
%\end{equation}
\end{itemize} 
For  $\BKsll$ matters are more complex which makes predictions in a first instance more complicated but in the long term allows to disentangle microscopic features of the interactions.
\begin{itemize}
\item \emph{Combination of Wilson coefficients} \\
Whereas $\BKll$ probes  $C_{9+}$, in $\BKsll$ both combinations $C_{9\pm}$ enter 
the decay rate, depending on the parity properties of the helicity amplitudes (e.g. \cite{HZ13}),
\begin{equation}
\label{eq:HH}
\HH_{\perp} \sim C_{9+}   \;, \quad \HH_{0,\parallel} \sim C_{9-}  \;.
\end{equation}
 We will parameterise the effect, extending the parameterisation \eqref{eq:hV}, as follows
\begin{equation}
C_9^{\rm eff} = (C_9 +   \ccm  \cnf h_c(q^2) + ... )  \;,\quad C_9^{'\rm eff} = (C'_9 +   \ccp  \cnf h_c(q^2) + ... ) \;, \quad \cc|_{K} \to  (\ccm + \ccp)_{K^*} \;.
\end{equation}
Hence with information from $\BKll$ only  there is ambiguity in predicting $\BKsll$. 
Conversely this ambiguity can be resolved with the aid of $\BKsll$-observables.
To get an idea of the qualitative nature of the effect we choose the following three scenarios:
\begin{equation}
\label{eq:scenarios}
(i) \; \cc \equiv ( \ccm , \ccp) =    -1.25(1,1)  \;, \quad (ii) \; \cc  =    -2.5(0,1) \;, \quad (iii)\;  \cc  =    -2.5(1,0)    \;.
\end{equation}
An important general strategy is to find observables which are sensitive to $C_{9-}$. 
A sizeable difference in  $ C_{9+} - C_{9-}  = 2C_{9'}  $  is a direct sign of the presence of right-handed currents and structure beyond the SM.

\begin{figure}[h!]
{ \includegraphics[scale=0.64]{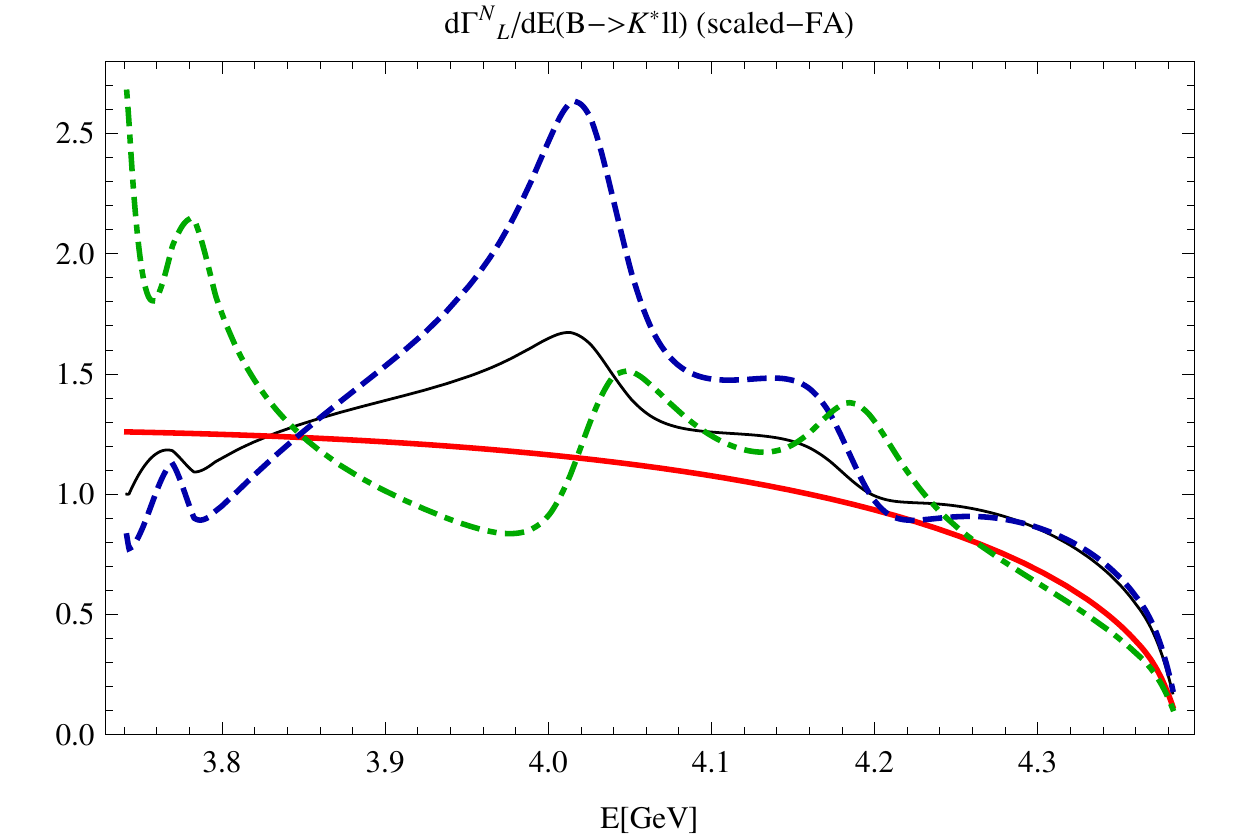} \;
 \includegraphics[scale=0.70]{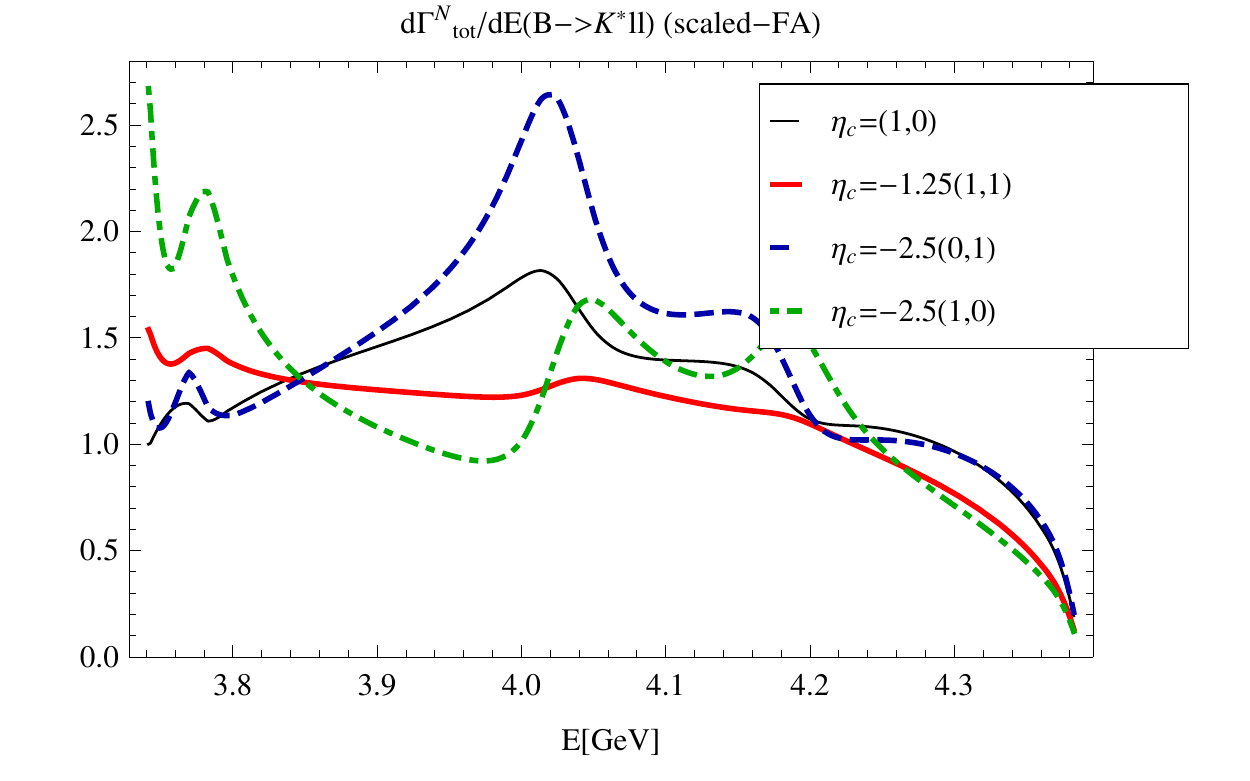}  
}
 \caption{\small  Plots of the longitudinal and total  decay rate  
 in the high $q^2$ region  for different values of  $\cc = (\ccm,\ccp)$ in the scaled-FA.
 The normalisation is such that $d\Gamma^N/dq^2(q^2 = 14 \GeV^2) = 1$  for $\cc = (1,0)$.
  The plots are useful to  distinguish between 
 the three scenarios \eqref{eq:scenarios}. Comments on the computation are the same as in the caption of Fig.\ref{fig:P5phigh}. The crossing of all four curves, just below the point $15\GeV^2$ originates from 
 $\Delta( C_9^{\rm eff} + C_9^{'\rm eff}) =0 $ going through zero at the same point c.f. Fig.~\ref{fig:C9eff}.
The real part interpolates between a dip and a peak through zero and the imaginary part goes to zero 
since the resonances $\Ra$ and $\Rb$ are spaced widely enough from each other.}
 \label{fig:Brhigh}
 \end{figure}

\begin{figure}[h!]
{ \includegraphics[scale=0.60]{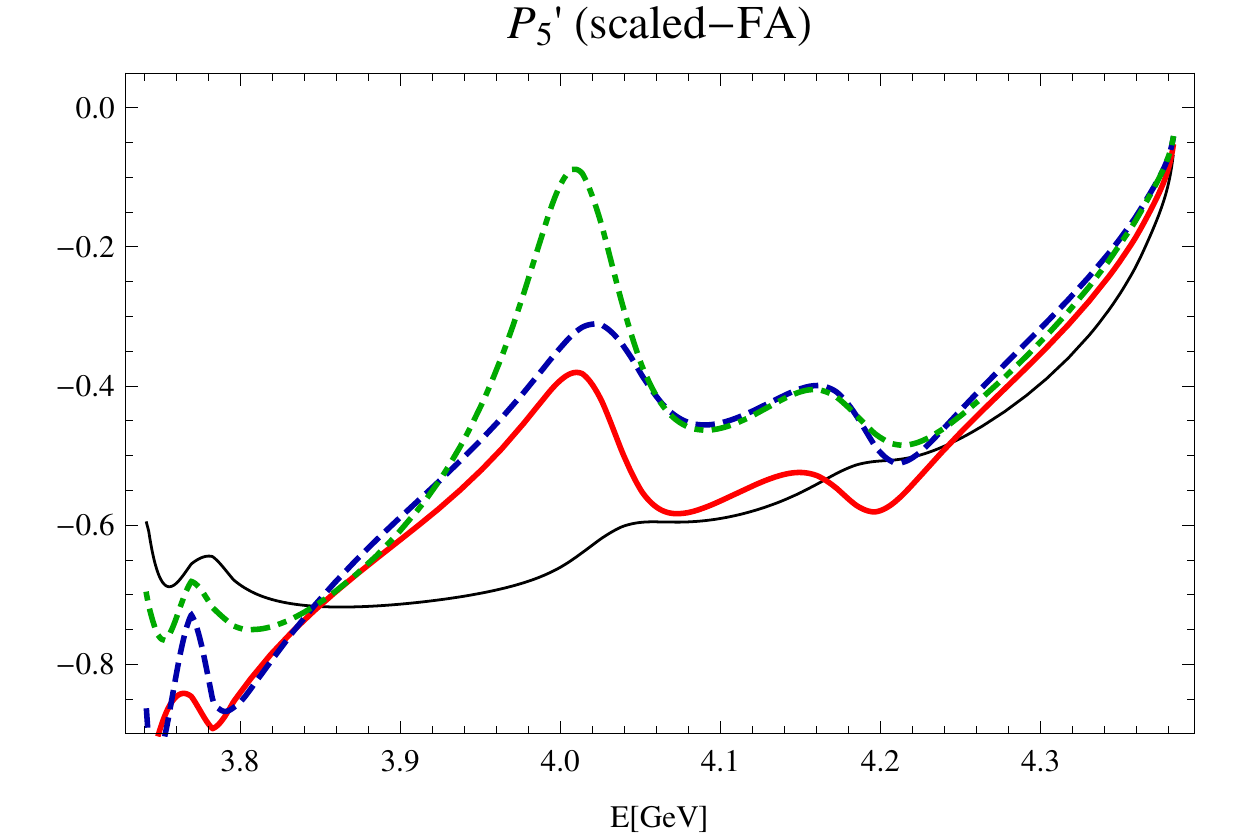} \;
 \includegraphics[scale=0.60]{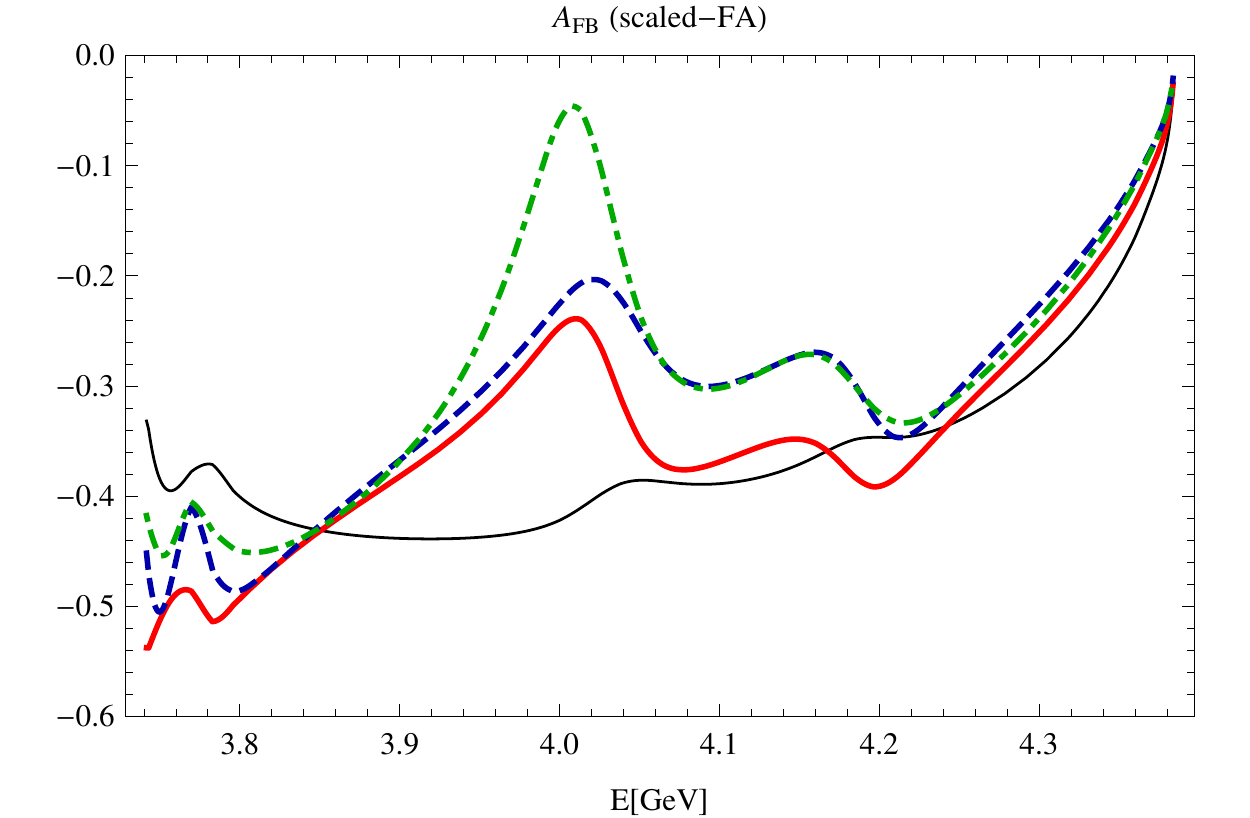}  \\[0.2cm] 
 \includegraphics[scale=0.60]{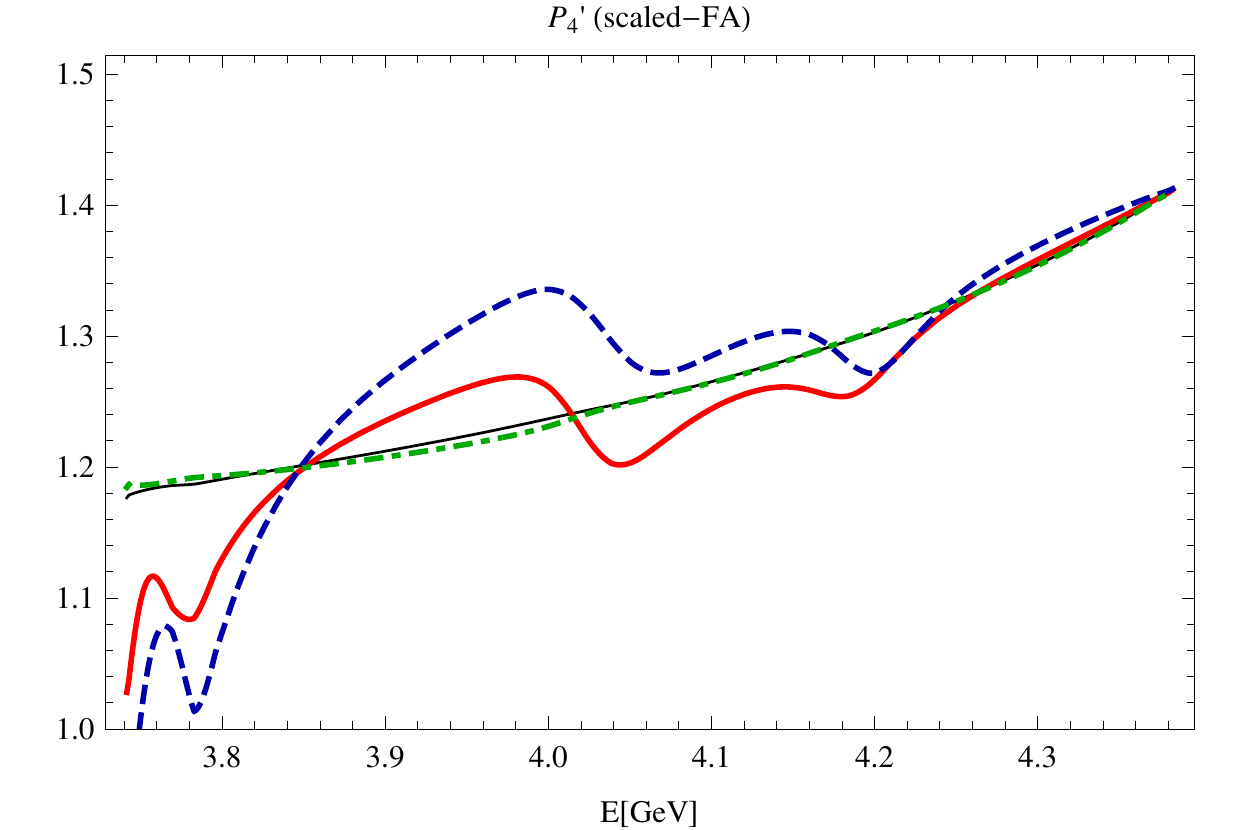} \;
 \includegraphics[scale=0.66]{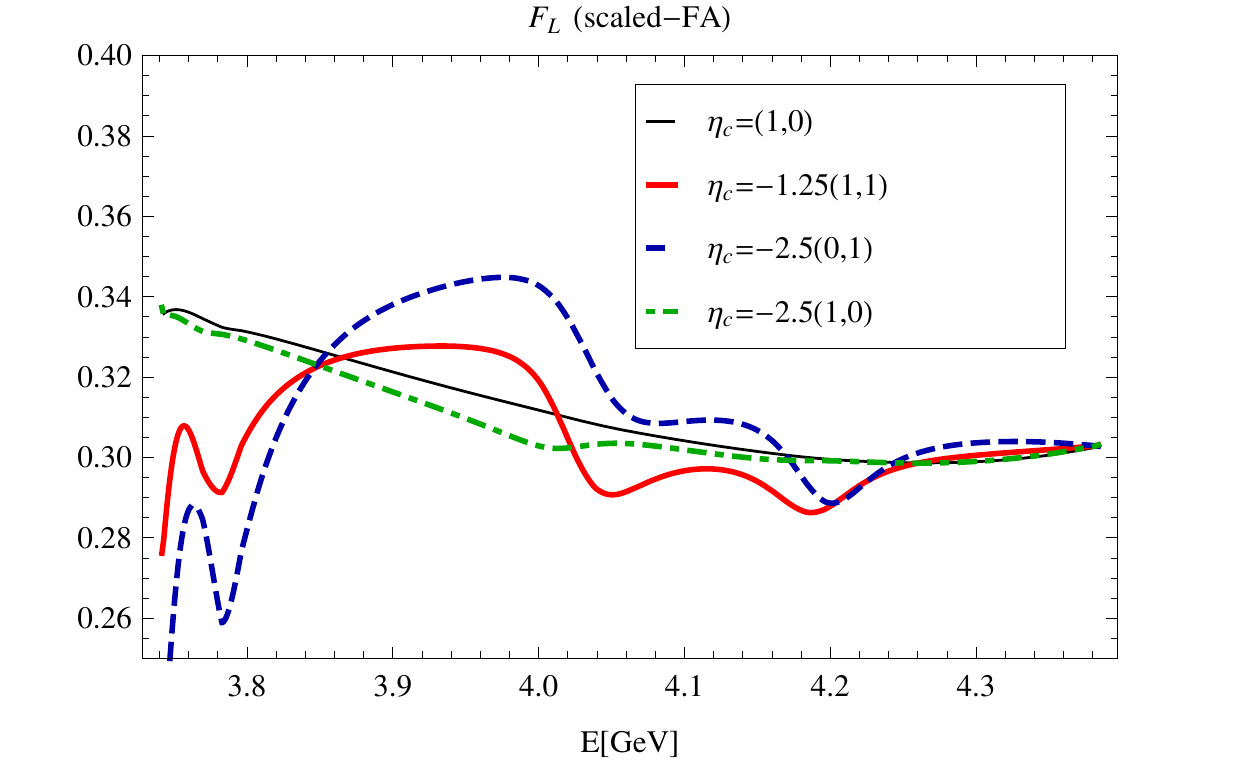} 
}
 \caption{\small  Plots of $P_5'$, $A_{\rm FB}$, $P_4'$ and $F_L$ for different values of  $\cc = (\ccm,\ccp)$ in the scaled-FA. The observables $P_4'$ and $F_L$, in the $\cc$-scaled FA , are insensitive to changes in $\ccm$ but sensitive to changes in $\ccp$ and therefore right-handed currents. The exact endpoint predictions \cite{HZ13} are $P_5'( \qmax) = 0$,  $A_{\rm FB}( \qmax) = 0$, $P_4'( \qmax) = \sqrt{2}$ and $F_L( \qmax) = 1/3$ (with $\qmax = (m_B - m_{K^*})^2$).  The similarity of $A_{\rm FB}$ and $P_5'$ is no accident since their ratio $P_5'/A_{\rm FB}(\qmax) = \sqrt{2}$.
 It is noted that $F_L( \qmax) \simeq 0.31$ in the actual plot and not $1/3$.
 This might be due to insufficient precision in digits of the lattice fits in \cite{latticeF} as well as the fact that 
 $A_{12} \to A_1$ at the endpoint is not exactly obeyed by the fits.
 The predictions are done using lattice form factors \cite{latticeF} in the high 
 $q^2$-region. Using the twist-3 ${\cal O}(\al_s)$ LCSR form factors \cite{BZ04b}, with updated values as in \cite{O8,HHSZ13}, we find that at $q^2 \simeq 14 \GeV^2$ the observables differ typically by about $3-4\%$ which is well below the uncertainties of both approaches. 
 No ${\cal O}(\as)$ vertex corrections are included since they are partly contained in the fit.}
 \label{fig:P5phigh}
 \end{figure}

\item \emph{The high $q^2$ (low recoil) region}
\begin{itemize}
\item At high $q^2$ ($q^2 > 14 \GeV^2$)\footnote{It would be desirable if LHCb would release data on the narrow charm-resonances $J/\Psi$ and $\Psi(2S)$. One could imagine  to use the information  in many ways.}  there is information on the behaviour of the  charm-resonances and predictions seem most promising in this region.  
%The transformation of this information to the low $q^2$-region is necessarily subtler and dependent on the microscopic model. We have devoted subsection \ref{sec:lowq2} to this topic. 
 At the very endpoint the values of the angular observables, for the 
effective Hamiltonian \eqref{eq:SMbasis}, are exact and follow from Lorentz-covariance only \cite{HZ13}; independent of  approximations and values of the Wilson coefficients.  
For observables with finite value at the endpoint, LHCb-data  in the last bin is, fortunately, 
found to be in agreement 
(c.f. table II \cite{HZ13}) with the estimated average deviations of around $10$ - $15 \%$.\footnote{
The slope of observables vanishing  linearly in the momentum at the endpoint (e.g. $A_{\rm FB}$, $P_5'$) carry a degree of universality. One can either build ratios which follow exact prediction or fit for the slope which is sensitive to new physics \cite{HZ13}.}\footnote{\label{foot:P4p} The value $\aver{P_4'}_{[14.18,16]\GeV^2} = -0.18^{+0.54}_{-0.70}$ therefore seems rather far of from 
the endpoint value  $\aver{P_4'}(\qmax)$ and it is therefore likely that this value will shift with the 
\threefb-data. The notation $\aver{P_4'}$ corresponds to  a bin averaging  explained in  appendix \ref{app:angular}. We wish to add that this procedure slightly  distorts 
the naive average from the plots in Figs.~\ref{fig:Brhigh},\ref{fig:P5phigh}.} 
The reader is referred to the plots in Fig.~\ref{fig:P5phigh} for illustration.
In essence, with further data this approach can extract valuable model-independent information from the endpoint region.
In the region away from the endpoint say $q^2  < 16-17\GeV^2$ the predictions deviate from their endpoint pattern 
and become increasingly sensitive to the scenario c.f. Figs.~\ref{fig:Brhigh},\ref{fig:P5phigh}.

\emph{Strategy 1:} the plots Figs.~\ref{fig:Brhigh},\ref{fig:P5phigh}  indicate that  the scenarios \eqref{eq:scenarios} can be determined from the total and longitudinal decay rate at a few $\GeV^2$ away from the endpoint.
% The fact that fit-c) and -d), to $\BKll$ in table \ref{tab:fit1}, look qualitatively 
%similar supports the believe that there is meaning of this statement beyond the $\cc$-scaled FA. 
\end{itemize}

\begin{itemize}
\item \emph{Polarisation dependent  non-factorisable contributions versus right-handed currents} \\
Factorisable corrections do factorise into a charm-loop part and a form factor. 
In this case the polarisation dependence is solely encoded in the form factor and 
therefore the same as the leading contribution as emphasised and used in the appendix of  \cite{HZ13}.  Under the assumption of the absence of right-handed currents charm-loop contribution 
(in  the FA  and $m_\ell = 0$), drop out (c.f. section C.1 \cite{HZ13} for a more precise formulation) in observables  of the form 
\begin{equation}
\frac{\HH^L_i \HH^{L*}_j + \HH^R_i \HH^{R*}_j}{\HH^L_l \HH^{L*}_k + \HH^R_l \HH^{R*}_k } \;, \quad  i,j,k,l=\perp,||,0 \;.
\end{equation}     
 Examples are  the longitudinal polarisation fraction $F_L$, $P_2 \sim A_T^{(2)}$ 
 and $P_4'$.  Hence:
 \begin{itemize}
 \item \emph{Strategy 2a:} in the absence of right-handed currents the observables $F_L$, $P_2$ and $P_4'$ can be  used to \emph{test for polarisation non-universality} of the non-factorisable corrections. We emphasise whereas non-factorisable contributions can be non-universal they do not have to be. In a light-cone OPE approach non-universality enters through helicity dependence of the light-cone distribution  amplitudes.   
 \item \emph{Strategy 2b:} within the scaled-FA the observables $F_L$, $P_2$ and $P_4'$ can be used to test 
 for right-handed currents, i.e. $C'$ Wilson coefficients. This feature is illustrated in 
 Fig.~\ref{fig:P5phigh} for $F_L$  and $P_4'$. It is seen that the SM curve $ (\ccm,\ccp) = (1,0)$ is identical to $(\ccm,\ccp) = -2.5(1,0)$ but qualitatively different from 
  $(\ccm,\ccp) = -2.5(0,1)$.
\end{itemize}
Strategy 2 is not capable of disentangling right-handed currents from the potential non-universality of non factorisable corrections. Right-handed currents can though be tested 
 for in the $J_3$-angular variable (c.f. appendix \ref{app:angular} for the definition) or in 
 $B_s \to \phi \gamma$ \cite{MXZ08} for instance.
 \item \emph{Strategy 3:} The observable  
 $\Gamma^N_L(q^2)(B \to K^* \ell \ell) \sim (J_{1c} - J_{2c}/3) \sim |H_0^V|^2 + |H_0^A|^2$ only depends on $C_-$ and not $C_+$. Hence for a measurement of comparable 
 quality to the  $\BKll$-rate one can fit for both $C_{9-}^{\rm eff}$
 Alternatively  $C_{9-}^{\rm eff} $ (or $\ccm - \ccp$) can be obtained from $B \to K^*_0(1430) \ell \ell$ since $J^{P}(K^*_0) = 0^+$ is a scalar of opposite parity to the $K$-meson. 
 \end{itemize}
\end{itemize}

\subsection{Connections to the 2013 LHCb-anomalies in $\BKsll$ at \onefb}
\label{sec:lowq2}

The first set of measurement of angular observables at \onefb
\cite{LHCbfirst} turned out to be broadly consistent with the SM. 
A refined analysis of observables  \cite{LHCbanomaly}, with reduced form factor dependence, gave rise  deviations which received considerable attention \cite{understanding,AS13,latticepheno,DMVproc}. 
In particular  a $3.7\sigma$-deviation was observed 
in the observable $P_5'$ in the $q^2 = [4.30,8.68]\GeV^2$-bin. 
All of the global fits analyses \cite{understanding,AS13,latticepheno,DMVproc} at high and low $q^2$ 
find values  of  $-2 < \Delta C_9 < -0.5$.
The possibility of $\Delta C_9'  \simeq 1$, driven by high $q^2$ and $B \to K\mu\mu$,
 was suggested in \cite{AS13} and later by \cite{BG13,Beaujean:2013soa,latticepheno}.  
On the other hand at low $q^2$, and in particular for $P_5'$, $\Delta C_9'  \simeq -1$ \cite{DMVproc}.
As previously mentioned, inspection of Fig.~\ref{fig:C9eff}, makes it clear that the anomaly in the charm-resonances leads to qualitatively  similar effects. 
The crucial difference is though that  we interpret the effect 
as new $\bar b s \bar c c $ rather than $O_9^{(')}$-
operators since the latter do not give rise to the pronounced $q^2$-behaviour in the 
open charm-region.\footnote{An alternative possibility, allowed by the global fits 
\cite{Beaujean:2013soa}, is the flipped-sign solution where all the penguin Wilson coefficients flip sign 
$C_{7,9,10} \to - C_{7,9,10} $. This would give rise to a much better agreement of FA 
with data. Although the leading ${\cal O}(\al_s)$-corrections would still lower the effect in 
the opposite direction. In the language of $\cc$ we pass from 
$\cc \simeq (1-0.5)
 = 0.5 \to (-1+0.5) = -0.5$ which is definitely closer to $\cc = -2.5$ (fit-b) in table \ref{tab:fit1}) but still not close enough. The other problem is that we are not aware of a model or a mechanism that could give rise to 
 such a behaviour and we therefore discard this possibility  for the remaining part of this paper.}
 Our viewpoint is not compatible, in  a first instance, with the interpretation of the effects as coming 
 from a $Z'$-boson mediating between a $bs$ and $\ell\ell$-fields \cite{Zp,BG13}.
 
 For the computation at low $q^2$ we use the LCSR form factors \cite{BZ04b}, with updated values as in \cite{O8,HHSZ13}.  It is only for $\cc = (1,0)$ that we include the non-factorisable vertex-corrections from reference \cite{AACW01} as for the other cases this effect has been implicitly fitted for through the 
 scale factor $\cc$. 
   Hence an  inclusion of \cite{AACW01}  amounts to a degree of double counting which has to be avoided.
 
\begin{figure}[t]
{ \includegraphics[scale=0.66]{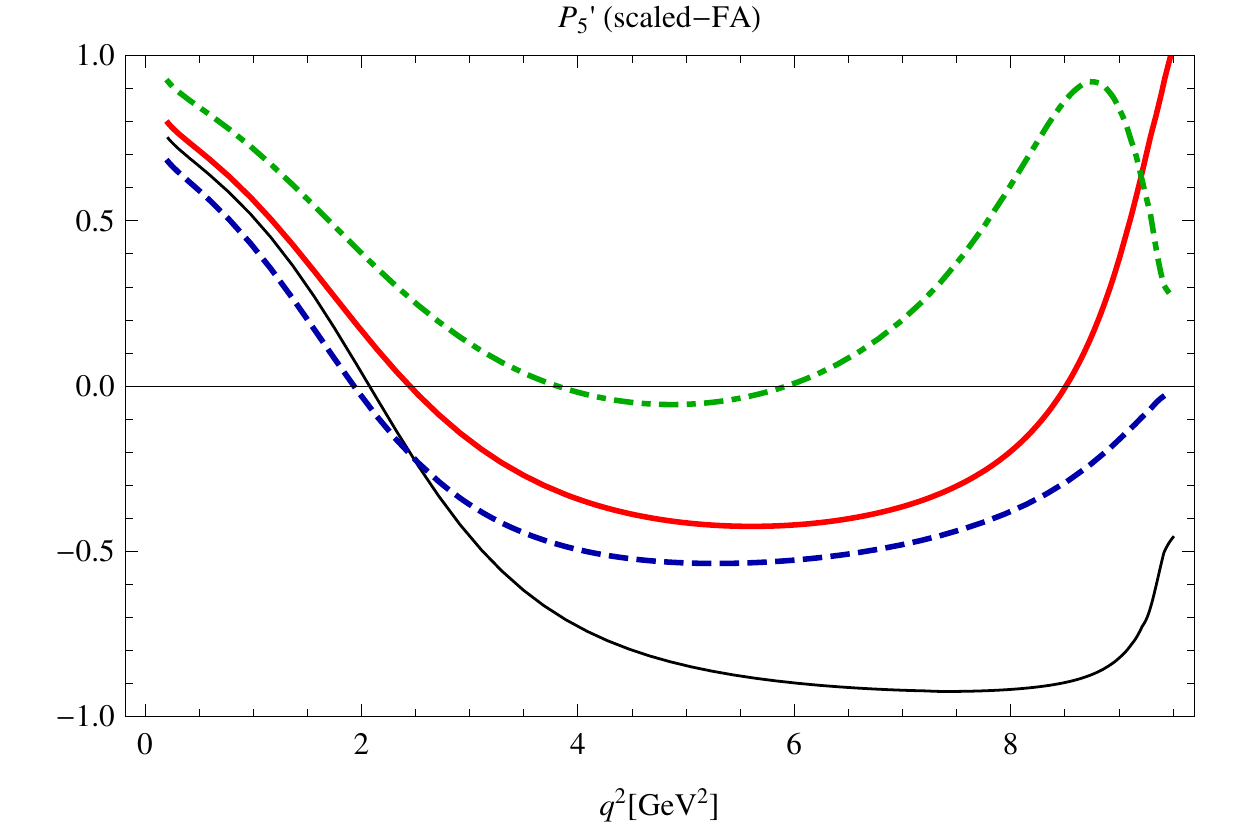} \;
 \includegraphics[scale=0.66]{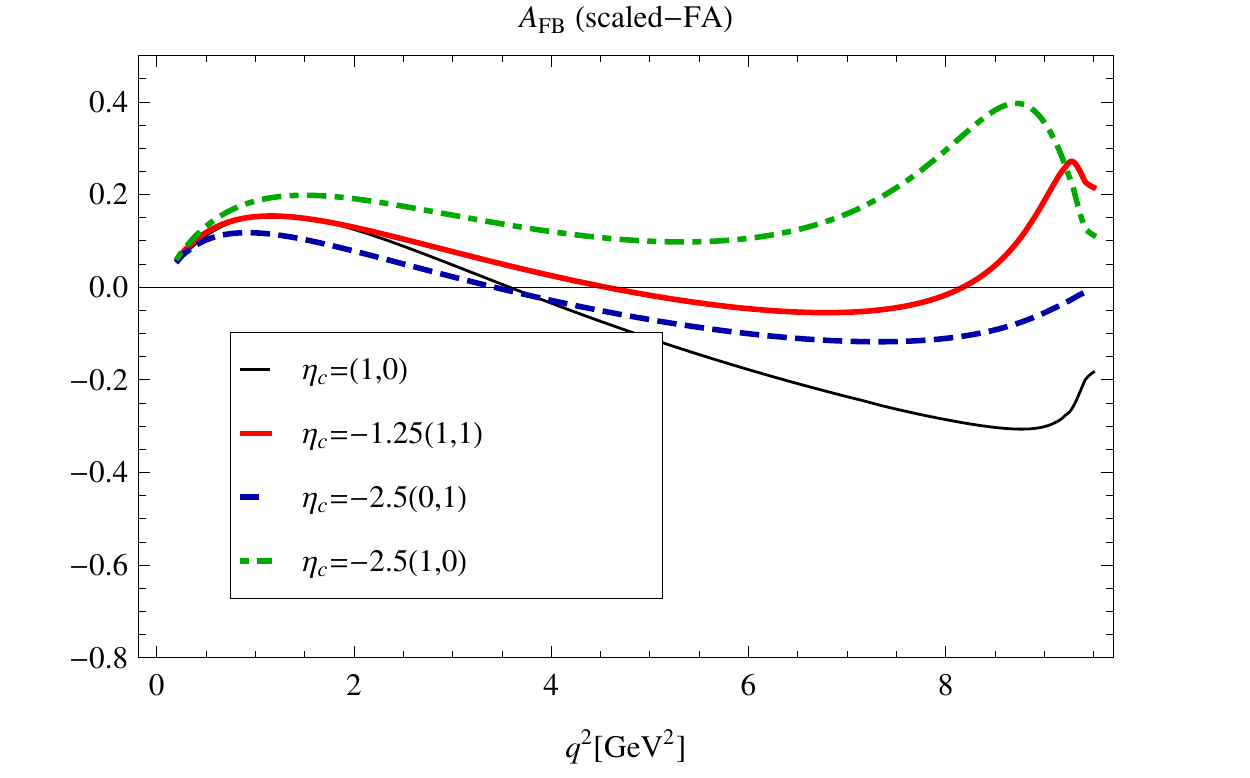}  \\
 \includegraphics[scale=0.66]{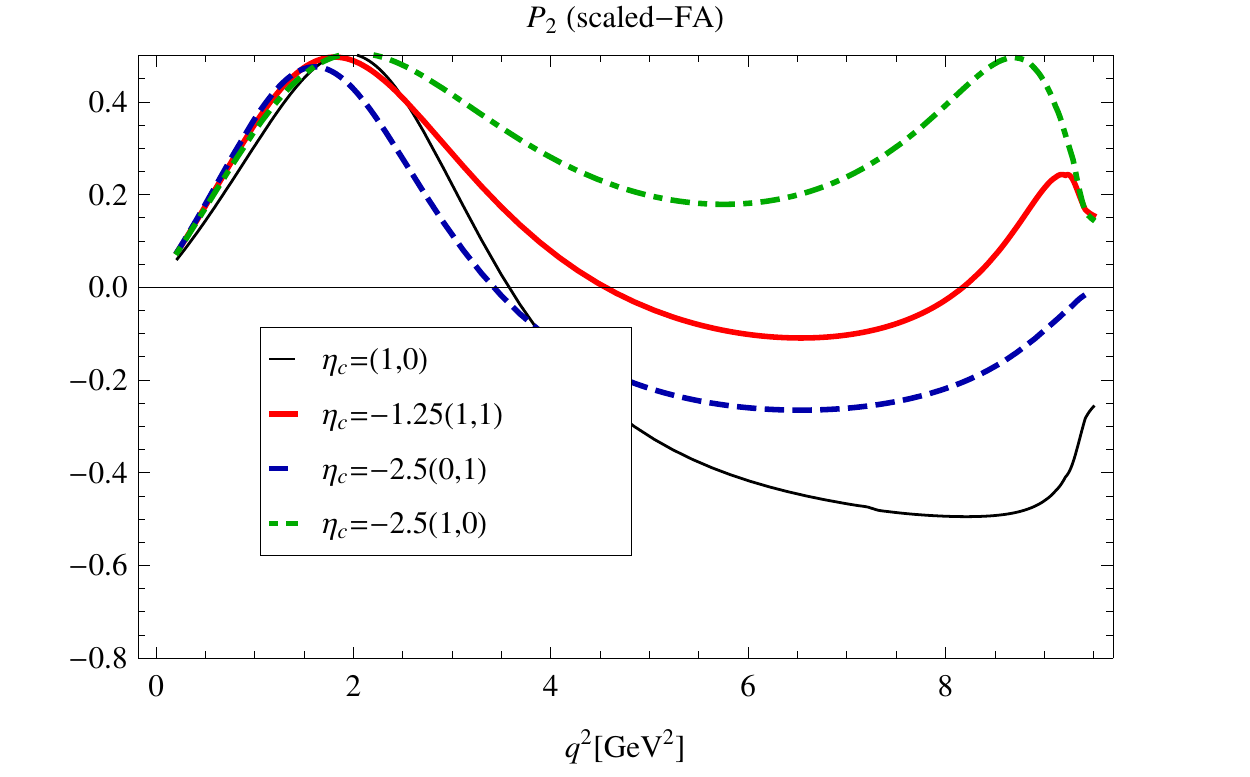} \;
 \includegraphics[scale=0.66]{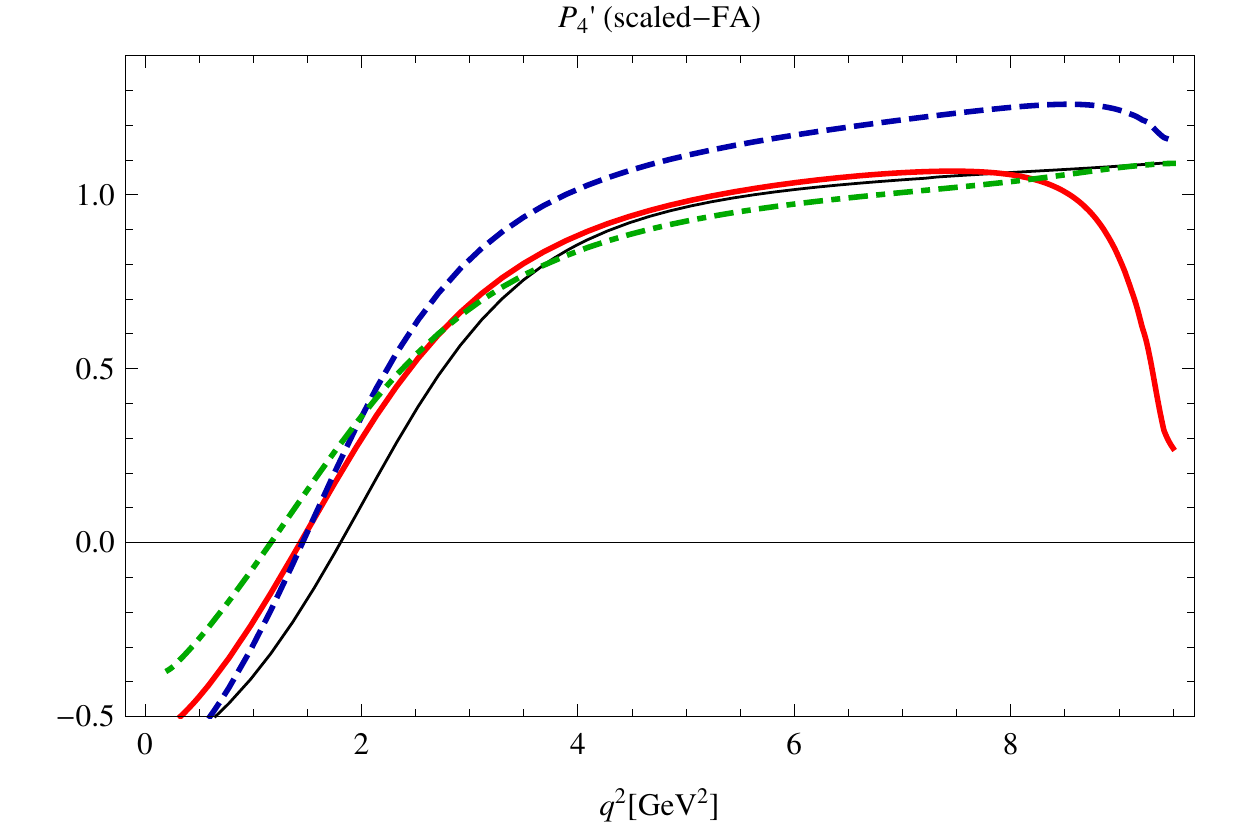}  
}
 \caption{\small  Plots for the same observables as in Fig.~\ref{fig:P5phigh}.  
 The form factors are taken from the LCSR computation \cite{BZ04b}, with updated 
 values as in \cite{O8,HHSZ13}.
 From the figures one can anticipate the changes in the binned observables shown 
 in table \ref{tab:Ps}.}
 \label{fig:P5pAFBlow}
 \end{figure}

 The plots are presented in Fig.~\ref{fig:P5pAFBlow} and binned observables are given in table \ref{tab:Ps}.  In particular the large deviations in $P_5'$ can be accounted for without making any of the predictions substantially worse.  
 We caution the reader that these numbers are meant for illustrative purposes only since there 
 is a degree of model-dependence through the inference from high to low $q^2$ in the absence of 
 a precise microscopic model.  As previously explained it is for this reason that we have confined ourselves
 to fit-b) for illustrating the effects.
  Yet the $\cc = (1,0)$-computation shown in black in Fig.~\ref{fig:P5pAFBlow} 
 is a good prediction in the SM-FA against which future experimental data can be discriminated against.
 
 Table \ref{tab:Ps} indicates that the data favours  scenario  $(i)$ with $\cc = -1.25$ and $\cc' = -1.25$. The only observable  which is significantly worse than in the SM-FA approximation is 
 $A_{\rm FB}$ in the $ [4.30,8.68] $-bin. Inspecting the table it would seem that a mixture of scenarios
 (i) and (iii) could give the best fit.\footnote{It would be interesting to fit the  $(\cc,\cc')$-pair to a complete set of $b \to s\ell \ell$-observables.} The experimental errors are  too large to draw solid conclusions at this point. 
 It is nevertheless worth to emphasise, once more, that the central value of this (mixed) scenario 
 would correspond to a sizeable $\Delta C_9'$-contribution which is a definite signal of new physics. 
  It seems unlikely, though possible, that such findings would be significantly changed  under a refinement of the fit model.

%\begin{figure}[h!]
% \includegraphics[scale=0.70]{angular_prefactor.pdf} 
% \caption{\small bla bla .}
% \label{fig:ang_pre}
% \end{figure}

  \begin{table}[ht]
\begin{center}
\begin{tabular}{c | c   rrrrr  }
Observable & $q^2$ & LHCb & SM & $\cc=$-$1.25(1,1)$  & -$2.5(0,1)$  & -$2.5(1,0)$  \\
\hline
%{0.0085, 0.16, -0.013, 0.33}, {0.15, 0.25, 0.067, 0.39}, {-0.44, \
%-0.05, -0.23, 0.29}, {-0.42, -0.39, -0.36, -0.36}, {-0.34, -0.31, \
%-0.25, -0.25}
%{0.57, 0.66, 0.8, 0.64}, {0.61, 0.69, 0.82, 0.67}, {1., 1., 1.2, 
 % 0.98}, {1.2, 1.2, 1.2, 1.2}, {1.3, 1.3, 1.3, 1.3}}
%-0.44, -0.15, -0.33, 0.17}, {-0.47, -0.17, -0.36, 0.13}, {-0.88, \
%-0.31, -0.44, 0.26}, {-0.7, -0.66, -0.59, -0.61}, {-0.53, -0.49, \
%-0.39, -0.38}
%{0.0026, 0.054, -0.0033, 0.14}, {0.034, 0.069, 0.014, 0.15}, {-0.21, \
%-0.025, -0.098, 0.19}, {-0.43, -0.4, -0.36, -0.37}, {-0.35, -0.33, \
%-0.26, -0.26}
$ \langle P_2 \rangle $ & $ [1.00,6.00] $ & $0.33^{+0.11}_{-0.12}$ & 0.0085& 0.16& -0.013& 0.33  \\
$ \langle P_2 \rangle $ & $ [2.00,4.30] $ & $0.50^{+0.00}_{-0.07}$ & 0.15& 0.25& 0.067& 0.39 \\
$ \langle P_2 \rangle $ & $ [4.30,8.68] $ &$-0.25^{+0.07}_{-0.08}$  & -0.44& -0.05& -0.23& 0.29 \\ \hline
$ \langle P_2 \rangle $ & $ [14.18,16.00] $ &$-0.50^ {+0.03}_{-0.00}$  &  -0.42& -0.39& -0.36& -0.36 \\
$ \langle P_2 \rangle $ & $ [16.00,19.00] $ & $-0.32^ {+0.08}_{-0.08}$ & -0.34& -0.31& \
-0.25& -0.25  \\ \hline \hline
$ \langle P_4' \rangle $ & $ [1.00,6.00] $ & $0.58^{+0.32}_{-0.36}$ &  0.57& 0.66& 0.8& 0.64  \\
$ \langle P_4' \rangle $ & $ [2.00,4.30] $ & $0.74^{+0.54}_{-0.60}$  & 0.61& 0.69& 0.82& 0.67 \\
$ \langle P_4' \rangle $ & $ [4.30,8.68] $ & $1.18^{+0.26}_{-0.32}$ & 1.0& 1.0 & 1.2& 
  0.98 \\  \hline
$ \langle P_4' \rangle $ & $ [14.18,16.00] $ & $-0.18^ {+0.54}_{-0.70}$ & 1.2& 1.2& 1.2& 1.2 \\
$ \langle P_4' \rangle $ & $ [16.00,19.00] $ &  $0.70^ {+0.44}_{-0.52}$ & 1.3& 1.3& 1.3& 1.3 \\ \hline \hline
$ \langle P_5' \rangle $ & $ [1.00,6.00] $ & $0.21^{+0.20}_{-0.21}$ & -0.44& -0.15& -0.33&
 0.17 \\
$ \langle P_5' \rangle $ & $ [2.00,4.30] $ & $0.29^{+0.40}_{-0.39}$  & -0.47& -0.17& -0.36& 0.13 \\
$ \langle P_5' \rangle $ & $ [4.30,8.68] $ & $-0.19^{+0.16}_{-0.16}$ & -0.88& -0.31& -0.44& 0.26 \\ \hline
$ \langle P_5' \rangle $ & $ [14.18,16.00] $ & $-0.79^ {+0.27}_{-0.22}$ & -0.7& -0.66& -0.59& -0.61 \\
$ \langle P_5' \rangle $ & $ [16.00,19.00] $ &$-0.60^ {+0.21}_{-0.18}$ & -0.53& -0.49& 
-0.39& -0.38 \\ \hline \hline
$ \langle A_{\mathrm{FB}} \rangle $ & $ [1.00,6.00] $ & $0.17_{-0.06}^{+0.06}$ & 0.0026& 0.054& -0.0033& 0.14 \\
$ \langle A_{\mathrm{FB}} \rangle $ & $ [2.00,4.30] $ &$0.20^{+0.08}_{-0.08}$ &0.034& 0.069& 0.014& 0.15 \\
$ \langle A_{\mathrm{FB}} \rangle $ & $ [4.30,8.68] $ & $-0.16^{+0.05}_{-0.06}$ & -0.21& -0.025& -0.098& 0.19 \\  \hline
$ \langle A_{\mathrm{FB}} \rangle $ & $ [14.18,16.00] $ & $-0.51^{+0.05}_{-0.07}$  &-0.43& -0.40& -0.36& -0.37 \\
$ \langle A_{\mathrm{FB}} \rangle $ & $ [16.00,19.00] $ &$-0.30_{-0.08}^{+0.08}$ &  -0.35& -0.33& -0.26& -0.26
\end{tabular}
\caption{\small From  left to right: LHCb-data,  SM prediction and the three scenarios \eqref{eq:scenarios} for the four observables shown Fig.~\ref{fig:P5pAFBlow}. The averaging procedure is described in appendix \ref{app:angular}. The result are given to two significant digits. We plan to give theory errors for the  SM prediction in an updated version of this paper.  Uncertainties are especially relevant in the 
$[1,6]\GeV^2$-bin since many observables cross zero throughout this bin.  
Disregarding this bin we see that agreement with the experimental data is generically improved. 
In particular the $3.7\sigma$ deviation of $P_5'$ in the  $q^2 = [4.30,8.68]\GeV^2$-bin is much improved. We have commented on the high $q^2$ issue of $P_4'$ in a previous footnote. 
An interesting possibility for future investigations is to fit for the parameters $\ccm$ and $\ccp$. 
Albeit see main text for comments.}
\label{tab:Ps}
\end{center}
\end{table}

\subsection{Brief discussion on origin and consequences  of new $\bar  b s \bar cc$-structures}
\label{sec:newOP}

 The  SM is a very successful theory in the sense that 
 it passes many non-trivial tests. The addition of new structure is generally highly constrained.
 We intend to briefly discuss to what extent new operators of the form 
 \begin{equation}
 \label{eq:typeccbs}
 \Op_{\Gamma_1 \Gamma_2} =  \bar c \Gamma_1 c  \bar b \Gamma_2 s \;,
 \end{equation}
 are interesting and possibly constrained. The symbols $\Gamma_{1,2}$ stand for 
 Dirac matrices or covariant derivatives in which case the operators are of higher dimension 
 than the minimal four quark operators. We shall not discuss structures with colour since they 
 do not contribute to the FA and are  therefore generically $\al_s/(4 \pi)$-suppressed. 
 We collect a few observations below: 
 \begin{itemize}
 \item Generically we expect the (CP-odd) weak phases of the operators close to the  SM one since many of them 
 would affect the extraction of the CKM angle $\sin (2 \beta)$ 
  through decays like $B \to J/\Psi K_s$.
 \item 
  The current-current Wilson coefficient $C_2$ 
 mixes into the penguin operators in a significant way. 
For example  the  agreement of $b \to s \gamma$ between experiment and theory (within errors) 
implies that  $C_7^{\rm eff}(m_b)$ has to be close to its SM-value.  
More precisely since  $C_7^{\rm eff}(m_b)|_{\rm SM} \simeq -0.3   \simeq -0.14  - 0.16 C_2(M_W)$, 
$\Delta C_2(M_W)  \ll C_2(M_W)=1$ is natural in the absence of systematic cancellations. 
Hence $\cc$ seems contrived from the viewpoint of electroweak-scale new physics.
 \item Ignoring this aspect\footnote{New structure might, for example, be related to a non-minimal dark matter sector with non-trivial flavour structure.} there are are still  constraints from 
 the $b$-quark scale to be accounted for. 
 Operators of the type  \eqref{eq:typeccbs} potentially contribute to $\Delta \Gamma_s$  through  closed charm-loops. 
 These observables are highly constrained by current data yet 
  at ${\cal O}(\al_s^0)$ the contribution  can be avoided if $\matel{0}{b \Gamma_2 s }{B_s} = 0$ which is the case for all Dirac structures $\Gamma_2$ except  $\gamma_5$ and $\gamma_\mu \gamma_5$. 
  We refer the reader to reference \cite{BsmixingBSM} for a discussion of effects 
  of new physics on $\Delta \Gamma_s$. 
 In a low scale new physics scenario the study 
  of higher dimensional operators seems imperative  since the $m_b/m_W$-suppression argument does not apply.  
  The renormalisation group evolution and  classification of dimension 7 operators 
  for $b \to s \ell \ell$ has been studied in \cite{Karls}.
  \item A very important aspect is that for $\Gamma_1 \neq \ga_\mu$ the charm-loop in the FA is not described by a  diagonal correlation function. This has two consequences: (i) the residues $r_r$ in \eqref{eq:BW} are not proportional to the residues in $e^+e^- \to \text{hadrons}$ (ii) the residues $r_r$ are generally  not positive (note that (ii) implies (i)). 
   Or in terms of a model independent statement: the discontinuity is not positive definite anymore. Hence the tendencies, although not compelling, seen in fits c) and d) in table \ref{tab:fit1} 
   for scaled residues might be explained by \emph{either} the presence of new operators in the FA or sizeable non-factorisable corrections. 
  \item The quantitative description of $B \to  (\bar cc) K^{(*)}$-decays has a long and problematic history.
 %   The description of decays of the type $B \to J/\Psi K$  falls short in naive facorisation and it is presumed that there are large non-factorisable corrections.
  The situation is most pronounced for the $P$-wave charmonium states $\chi_c$. 
 For example  ${\cal B}(B \to \chi_{c0} K) = 1.47(27) \cdot 10^{-4}$,  ${\cal B}(B \to \chi_{c1} K) =   3.93(27) \cdot 10^{-4}$ \cite{PDG} where the former but not the latter vanishes in the FA.  In the SM it is usually concluded that 
 there have to be large non-factorisable corrections.
   In QCD factorisation  non-factorisable corrections, which were shown to be free of endpoint divergences upon inclusion of colour octet operators \cite{BV08},  
 lead to a qualitative improvement in many aspects. 
 On the quantitative level  the essence of the analysis  \cite{BV08} (c.f. figure 9 in that reference) 
 seems to be that for very large charm masses     
 the $\chi_{c0,c1}$ branching fractions can nearly be accommodated for but the smallness of  
 ${\cal B}(B \to \chi_{c2} K) < 1.5 \cdot 10^{-5}$ \cite{PDG} remains unresolved. With the current PDG numbers the mismatch is roughly a factor of five or higher.
Even in the absence of a concrete approach to non-facorisable correction it seems 
 difficult to explain why $\chi_{c2}$-rate is so small as compared to the $\chi_{c0}$-rate when 
 both vanish in the FA.
 It is tempting to speculate  that this puzzle is related to our findings. It might be possible to introduce operators which contribute at ${\cal O}(\al_s^0)$ to $B \to \chi_0 K$ but not to $B \to \chi_2K $.
 \end{itemize}

  \subsection{Summary of consequences and strategies}
  \label{sec:summary-end}
  
     In this subsection we would like to address possible improvements and further  steps of investigation. 
   Without further data  the basic directions  are to analyse  whether or not QCD can explain 
   the excess and to investigate the effects of  operators of the type \eqref{eq:typeccbs} on all kinds of observables through computations and global fits. 
   With the advent of new data there are, as usual, new possibilities.
   First, one can try to fit for $C_9-$ in the high $q^2$-region as outlined above. 
   Second, deviations in the low $q^2$-region below the $J/\Psi$-resonance, where  perturbation theory is trustworthy,  are signals whose effects cannot be associated with resonance physics.  
   Yet,  based on our findings, we expect deviations to grow towards the $J/\Psi$ resonance region.  
 Third, one could improve  on the  fit-model by  a K-matrix formalism, including information on the 
 $J/\Psi$, $\Psi(2S)$ and incorporating  a background model for the discontinuity. 
 Then one can use the dispersion relation \eqref{eq:d2} to obtain the amplitude and fit the real subtraction constant from the data.
 %Whether or not the high $q^2$-data is good enough to fit a subtraction constant is a non-trivial and interesting question that could be explored in future work.

\section{Summary and conclusions}
\label{sec:discussion}

We investigated the interference effect of the open charm-resonances with the short distance penguins in the SM.  The interference seen in the LHCb-data \cite{LHCb13_resonances}
shows a more pronounced structure with opposite sign as compared to the  FA 
of the SM. The FA prediction follows from first principles from a dispersion relation through 
$e^+e^- \to \text{hadrons}$ (BESII-data). 
Whether or not the effect can be described within the SM depends on the size of non-factorisable correction. 
The latter are $\al_s$-suppressed but colour enhanced. A parton estimate indicates an average correction 
of a factor $\sim -0.5$ as compared to the FA which is too small a result by a factor of seven.  By performing combined 
fits to the BESII- and LHCb-data we  tested for cancellations under the duality-integral. 
We found no indications that the parton estimate falls shorts by  a sizeable amount. 
In this first analysis, we have not found any signs that the SM or QCD respectively could account for the effect. Physics related to  charm is known to be a notoriously difficult  and 
further investigations are certainly highly desirable.

We have shown that the effect is presumably connected to the $\BKsll$-anomalies found in 2013. 
Out of the three scenarios chosen, to resolve the ambiguity from passing to $C_{9+}\equiv C_9 + C_9'$ to $C_{9-}\equiv C_9 - C_9'$, the $\BKsll$-data favours
  scenario (i) with $ \Delta C_9 = \Delta C_9'$. The reader is referred to  table \ref{tab:Ps} and  section \ref{sec:lowq2} for further remarks. 
  We have devised strategies to test for the microscopic structure of the effect.
For example the $\BKsll$ and the $B \to K^*_0(1430)\ell\ell$ observables are sensitive to the opposite parity combination $C_{9-}$  of Wilson coefficients. 
The knowledge of both parity combinations would allow to infer on right-handed currents which cannot be explained by QCD interactions.  
In the last  section we have given a brief outlook on consequences of the effect and how they could relate to 
other observables and old standing puzzles such as the non-leptonic  $B \to (\bar cc) K^{(*)}$-decays.

One of the most important outcomes of our investigations are that 
the anomalous resonance-behaviour, the 2013 $\BKsll$-anomalies and 
presumably the $B \to (\bar cc) K^{(*)}$-decays   have the same roots. Whether it is new physics or aspects 
of strong interactions which we do not understand is the real question. 
%It would seem that the duality intervals over the open charm regions are less delicate than 
%non-leptonic $B \to (\bar cc) K^{(*)}$ decays. 
We therefore feel that these new findings sharpen the quest for investigations into $b \to s \bar cc$-physics.

\section*{Acknowledgement}
R.Z. is grateful for partial support by an advanced STFC-fellowship. 
We are grateful to the BES-collaboration, 
Ikaros Bigi, Martin Beneke, Christoph Bobeth, Greig Cowan, Christine Davies, Danny van Dyk, Ulrik Egede, Tony Kennedy,   Einan Gardi, Christoph Greub, Gudrun Hiller, Mikolai Misiak
Franz Muheim, Matt Needham,  Patrick Owen, Stefan Meinel  Mitesh Patel, Kostas Petridis, Steve Playfer, Nicola Serra, Christpher Smith, David Straub, Misha Voloshin as well as 
many of the participants of the $\BKsll$-workshop at Imperial College from 1-3 April 2014
for discussion, correspondence and alike.  We are grateful to James Gratrex for comments on the manuscript and David Straub for partial numerical crosschecks.  This work was finalised during pleasant and fruitful stays at  the FPCP-conference in Marseille and the $b \to s\ell \ell$-workshop in Paris from the 2nd-3rd of June 2014.
A more elaborate list of references will be added in an update.

\appendix
\numberwithin{equation}{section}

\section{Details of computation}

\subsection{The $\BKll$-decay rate}

The $\BKll$ rate  extended from \cite{LZ13} to include a transversal amplitude $\HH^t$ 
is given by
\begin{eqnarray}
\label{eq:dGdq2}
& & \frac{d\Gamma}{dq^2}^{B\to K \ell^+\ell^-} =  
 \left[  \frac{c_F  \lambda_{K}^{3/2} \beta_{l}^{1/2}   }{2 (m_B+m_K)^2 } \right] 
 \left(\frac{\alpha}{4\pi}\right)^2 \,   \Big[ \frac{1 +\beta_l^2/3}{2}( |\HH^V|^2 + |\HH^A|^2) + \frac{2 m_l^2}{q^2} 
 ( |\HH^V|^2 - |\HH^A|^2) + |\HH^t|^2 \Big]    \;,
\end{eqnarray}
where $c_F \equiv  (G_F^2 |\lambda_t|^2 m_b^2 m_B^3/12\pi^3)$ and  
the K\"all\'{e}n-functions with normalised entries are
\begin{eqnarray}
\lambda_{K} \equiv \lambda(1,  m_K^2/m_B^2, q^2/m_B^2) \;, \quad \beta_{l} \equiv \lambda(1,m_l^2/q^2,m_l^2/q^2)  = 1 - \frac{4 m_l^2}{q^2}  \;.
\end{eqnarray}
with $\lambda(x,y,z) \equiv  ((x+y)^2 - z^2) ((x-y)^2 - z^2)$.

With regard to the notation in \cite{LZ13} we $h_T^{V,A} \to \HH^{V,A}$ to lighten the notation and avoid confusion.
The  helicity amplitude with axial coupling to the leptons is 
\begin{alignat}{1}
\HH^{A}(q^2) = C_{10} \frac{m_B+m_K}{2m_b} f_+(q^2) ,
\end{alignat}
the transversal amplitude squared is given by\footnote{The transversal amplitude is suppressed by $m_l^2$ but is formally leading very close to the endpoint since $\la_K \to 0$ at the endpoint $q^2 = (m_B-m_K)^2$. The actual numerical impact is rather small: 
at  $\qmax - 0.1 \GeV^2$  the effect is about $1\%$ only on the rate. More interestingly $\BKll$ provides an opportunity to search for scalar operators $\bar b (\ga_5) s \bar \ell (\ga_5) \ell$ near the kinematic endpoint since they are not $m_l^2$ suppressed but $1/\la_K$-enhanced as can be inferred from the formulae in section IV.A \cite{HZ13}.}
\begin{equation}
|\HH^t|^2 = |C_{10}  
 \Big( \frac{  m_l^2  (m_B+m_K)^4 (m_B-m_K)^2}{\la_K q^2 m_b^2}   \Big)  f_0(q^2)|^2
\end{equation}
and the vectorial coupling is split into two parts $\HH^V = \HH^{V,0} + \HH^{V,q} $ 
\begin{alignat}{1}
\HH^{V,0}(q^2) &= C_9^{\mathrm{eff}}(q^2)\frac{m_B+m_K}{2m_b} f_+(q^2) + C_7^{\mathrm{eff}} f_T(q^2) \;, 
\nonumber  \\ 
 \HH^{V,q}(q^2) &= C_8^{\mathrm{eff}} G^q(q^2) + W^q(q^2) + S^q(q^2) \;,
\end{alignat}
where $\HH^{V,0}$ are the numerically relevant ones for the rate.  
The  function $f_{+,T}(q^2)$ are the standard form factors are defined later on.
The contributions to $\HH^{V,q}$ are the chromomagnetic $\Op_8$ matrix elements $G^q$ \cite{O8}, weak annihilation $W^q$
and the spectator quark correction $S^q$ \cite{LZ13} which are important for the isospin asymmetries, in part because they depend on the spectator quark $q = u,d$. 
Their contribution to the SM-rate is rather small and can be neglected in a first assessment. 
The chirality-flipped operators ${\cal O'}$ for $\BKll$ are included by replacing 
\begin{equation}
C_{..7,8,9,10} \to C_{..7,8,9,10+}  \equiv C_{..7,8,9,10} + C_{..7,8,9,10}'
\end{equation}
in the equations above by virtue  of parity conservation of QCD. 
Hence $\BKll$ only constrains $C_+$ Wilson coefficients.

The standard form factors, using the notation \cite{HHSZ13},  are given by
\begin{alignat}{2}
\label{eq:ff}
 & \matel{K(p)}{\bar s i q_\nu \sigma^{\mu\nu}  b}{\bar B(p_B)} \; &=& \;  P_T^\mu \, f_T(q^2)     \; ,\nonumber \\[0.1cm]
 & \matel{K(p)}{\bar s \gamma^\mu  b}{\bar B(p_B)} \; &=& \;  P_T^\mu \, v_T  +   q^\mu    \frac{m_B^2-m_K^2}{q^2} f_0(q^2)   \; ,
 %\nonumber \\[0.1cm]
%& \matel{K(p)}{ (2 i \!\stackrel{\leftarrow}{D})^{\mu}}{\bar B(p_B)} \; &=& \;  P_T^\mu \, {\cal D}_T(q^2)  +   q^\mu    {\cal D}_s(q^2) \;,
\end{alignat}
with projector $P_T^\mu  =  \{(m_B^2-m_K^2) q^{\mu} - q^2 (p+p_B)^\mu\} /(m_B+m_K) $ 
and $v_s$ and $v_T$ are given by:
\begin{equation}
 v_s =  \frac{m_B^2-m_K^2}{q^2} f_0(q^2)  \;, \qquad  v_T =  \frac{ - (m_B+  m_K)}{q^2 }  \, f_+(q^2) \;.
\end{equation}
The effective Wilson coefficients read
\begin{eqnarray}
\label{eq:Ceff}
& & C_7^{\mathrm{eff}} = C_7 - \frac{4}{9}\CBBL_3 - \frac{4}{3}\CBBL_4 + \frac{1}{9}\CBBL_5 + \frac{1}{3}\CBBL_6 \;,  \quad  C_8^{\mathrm{eff}} = C_8 + \frac{4}{3}\CBBL_3 - \frac{1}{3}\CBBL_5 \;, \quad C_9^{\mathrm{eff}}(q^2) = C_9 + Y(q^2)  \;,
\end{eqnarray}
with 
\begin{equation}
\begin{split}
\label{eq:Y}
Y(q^2) =& \cnf h_c(q^2)
 - \frac{h_b(q^2)}{2}\left(4\CBBL_3 + 4\CBBL_4 + 3\CBBL_5 + \CBBL_6\right) \\
& - h_u(q^2)\left(\frac{\lambda_u}{\lambda_t}\left(3\CBBL_1+\CBBL_2\right) + \frac{1}{2}\left(\CBBL_3 + 3\CBBL_4\right)\right) + \frac{4}{27}\left(\CBBL_3 + 3\CBBL_4 + 8\CBBL_5\right)  \;;
\end{split}
\end{equation}
For the purpose of the discussion of this paper  we have split off
\begin{equation}
\label{eq:cnf}
\cnf = \left(-\frac{\lambda_c}{\lambda_t}\left(3\CBBL_1 + \CBBL_2\right) + 3\CBBL_3 + \CBBL_4 + 3\CBBL_5 + \CBBL_6\right) \;.
\end{equation}
 In the literature the following notation 
is frequently used: $\cnf = 3 a_{\rm eff}$. The function $h_f(q^2)^{(0)}$, to leading order in perturbation theory in naive dimensional regularisation  and $\overline{MS}$-scheme,  
is given by 
\begin{equation}
\label{eq:h0}
h_f^{(0)}(s)  =
%\stackrel{s \ll 4 m_f^2}{=} 
\frac{4}{9}\left(\frac{5}{3}- v^2 -  \ln\frac{m_f^2}{\mu^2} 
 \right)- \frac{4}{9} \,(3-v^2) \,|v| \,
\left\{
\begin{array}{l}
\,\arctan\displaystyle{\frac{1}{|v|}}
\qquad\quad s< 4 m_f^2 \\[0.4cm]
\, \frac{1}{2}(\ln\displaystyle{\frac{1+ v}{1-v}} - i \pi)
\quad s > 4m_f^2 \,,
\end{array}
\right.  
\end{equation}
with normalised $c$-quark momentum $v(s) \equiv \sqrt{1-4m_f^2/s}$.  

\subsection{Numerical input}
\label{app:input}

For the $B \to K$ form factor in the high $q^2$ range we use 
the recent lattice QCD predictions of HPQCD with staggered fermions \cite{Lattice13}.
The uncertainties are below $10\%$ in the relevant kinematic range $q^2 > \qref$.

We compute the Wilson coefficients in the basis \cite{Chetyrkin:1996vx}
and transform them
into the pseudo-BBL basis defined in \cite[eq. 79]{Beneke:2001at},
which is equivalent to the BBL basis at leading order in $\al_s$.
%The results of this procedure are given in table \ref{tbl:wilson-coefficient-example}.
The complete anomalous dimension matrix  to three loops is taken from 
\cite{Czakon:2006ss}, and the expressions for
the Wilson coefficients $C_i$ at the electroweak scale are taken from
\cite{Bobeth:1999mk} for $C_{1-6}$ and $C_{9,10}$ and \cite{Misiak:2004ew} for 
$C_{7,8}^{\mathrm{eff}}$.
These are always employed at $\mu=M_W$ to set the initial conditions for the RG flow;
that is to say the uncertainty owing to $\al_s$ terms at this scale is ignored,
although uncertainty of the masses of the $W$ boson and top quark is accounted for.
Since $\al_s(M_W)\approx 0.11$ is small this should have a negligible effect on the overall uncertainty of our calculations. The values of the Wilson coefficients are given 
in table 8 of \cite{LZ13}.

\subsection{Colour suppression}
\label{app:colour}

We briefly describe, in this little appendix, what is meant by colour suppression in the context 
of FA versus non-factorisable contributions.
Let us introduced the two operators which are convenient for our discussion,
\begin{equation}
\Op_{\bar c c}^{(1)} \equiv  (\bar c c)_{V-A}(\bar s b)_{V-A}  \;, \quad \Op_{\bar c c}^{(8)} \equiv  (\bar c T^ac)_{V-A}(\bar s T^ a b)_{V-A} \l,
\end{equation}
where we use the same notation as in \eqref{eq:SMbasis} and $T_a$ are the colour $SU(3)$ Lie Algrebra generators normalised as ${\rm tr}[T_a T_b ] = 1/2 \delta_{ab}$. 
These two operators are equivalent to the current-current four quark operator $\Op_2^c$ and its colour partner 
$\Op_1^c$
\begin{equation}
C_1 \Op_1^c + C_2 \Op_2^c = C_{\bar cc}^{(1)} \Op_{\bar c c}^{(1)} + C_{\bar cc}^{(8)} \Op_{\bar c c}^{(8)} \;.
\end{equation}
The Wilson coefficients in the BBL basis \cite{Buchalla:1995vs} and the CMM-basis \cite{Chetyrkin:1996vx}
read
\begin{alignat}{3}
\label{eq:18}
& C_{\bar cc}^{(1)} \;&=&\; C^{\rm BBL}_1 + \frac{1}{3}C^{\rm BBL}_2 \;&=&\; \frac{1}{3}\left(\frac{4}{3} C_2^{\rm CCM} + C_1^{\rm CCM} \right) \;, \nonumber  \\
& C_{\bar cc}^{(8)} \;&=&\; 2 C^{\rm BBL}_2 \;&=&\; \frac{1}{3}\left(- C_1^{\rm CCM} + 6 C_1^{\rm CCM} \right) 
\end{alignat}
In these formulae  we neglect contributions of the penguin four quark operators 
$\Op_{3-6}$ in both bases, which is not consistent but satisfactory for the purpose of illustration.
At the scale $\mu = m_b$, $C_{\bar cc}^{(1)} \simeq 0.2$ and $C_{\bar cc}^{(8)} \simeq 2$ and therefore 
colour suppression amounts to about an order of magnitude. With $\al_s(\mu)/(4 \pi) \simeq 0.02$ the canonic 
estimate for an NLO calculation which is not colour suppressed is then $20\%$ ($10 \cdot 0.02 = 0.2$).
Of course this should be taken as a rough estimate 
as in actual calculation this is  refined  where many diagrams might contribute and the complexity  of different scales might enter. Note, generally, only the ${\cal O}(\al_s^0)$ terms are colour suppressed since all 
higher order corrections couple to the colour octet operator.
%O_1^q &= (\bar s_i q_j)_{V-A}(\bar q_j b_i)_{V-A}

\subsection{Scale dependence of $x_D$}
\label{app:scale}

The factorisable contribution comes with the colour suppressed (c.f. previous subsection) contribution of Wilson 
coefficients $\cnf =  3 C_{\bar cc}^{(1)} + .. =  3(C_1 + C_2/3) + .. $ (\ref{eq:cnf},\ref{eq:18}) 
which is known to have a sizeable scale dependence due to cancellation effects.
The very approach of integrating out the charm quark due to $q^2 \gg4 m_c^2$ suggests
that one should use a large scale $\mu \simeq m_b$. In the approach of the high $q^2$ OPE the amplitude factorises into the charm-bubble times  a form factor (for the vertex) corrections and the low scale  $m_K$ is only present in the form factor.  
This is the picture of factorisation into UV and IR physics. It might in principle be that in a full computation effects of $m_K$ would be visible but since $m_K$ is put to zero in $H^{V,\cor}$ it would also seem that this effect cannot be estimated by varying 
$\mu$ to a very low scale.

Moreover, it is clear that by varying there is a certain trading between 
the factorisable and non-factorisable contributions.
The special circumstance that we have got $H^{V,\fac}$ from the data, which is more reliable and
 improvable with experimental data, suggests to choose the scale 
towards the direction where the factorisable contribution grows in relative size. This is 
the case towards a high scale. 

\begin{figure}[h!]
 \includegraphics[scale=0.75]{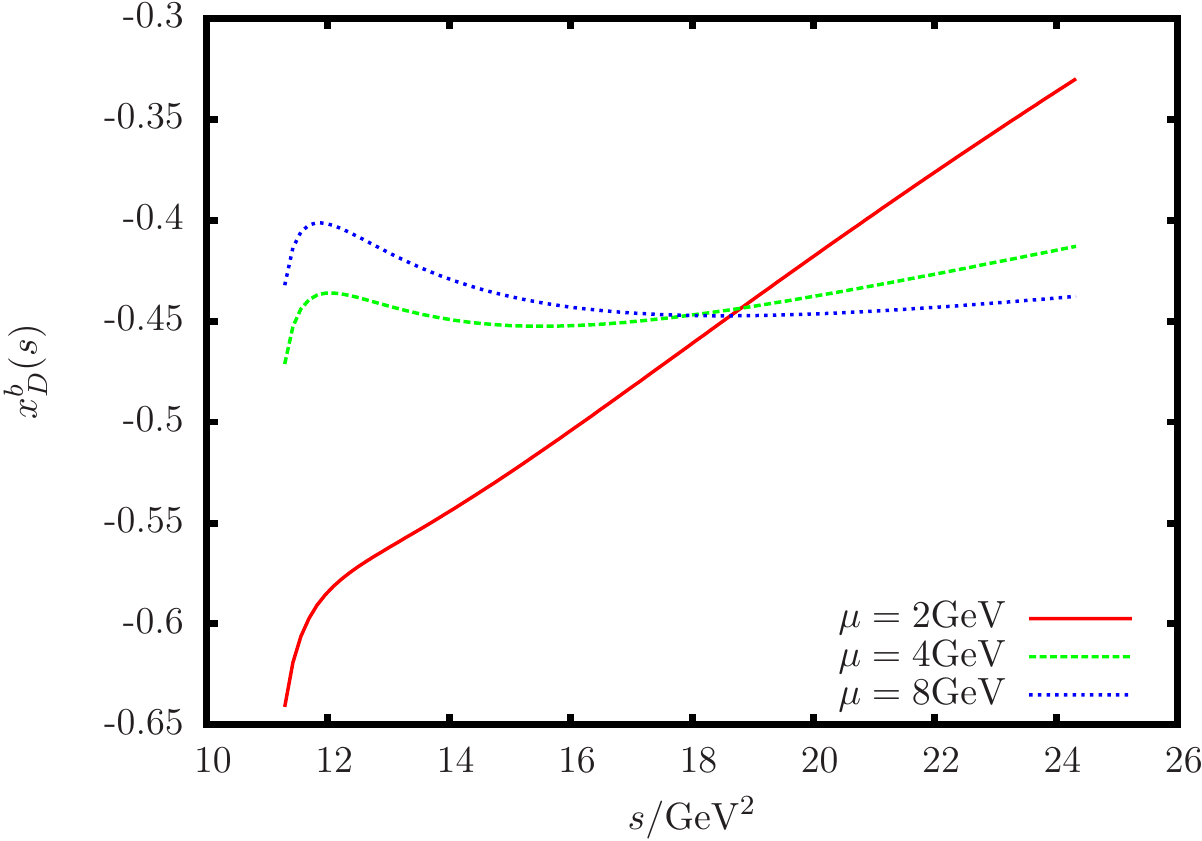} \\ 
 \caption{\small The function $x_D^b$ \eqref{eq:xDs} of vertex correction relative to facorisable corrections as a function of $q^2$ for three different reference scales $\mu = 2,4,8\GeV$. As argued in the main text a scale $\mu \simeq m_b$ is our preferred choice.}
 \label{fig:scale}
 \end{figure}

From these viewpoints it seems that $\mu > m_b$ rather than 
the other way around gives a reasonable number. 
The scale dependence of $x_D(q^2)$ \eqref{eq:xDs} is shown in Fig.~\ref{fig:scale} for a few reference scales aimed to help the reader to  form his or her own opinion.

\subsection{Angular observables in $B \to K^*\ell \ell$}
\label{app:angular}

A few of the angular observables used throughout the text with conventions as used 
in \cite{understanding}
  (except for reversed sign og $A_{\rm FB}$),
\begin{alignat}{4}
& P_1 &\; =\;&  \frac{J_3}{2 J_{2s}}  \;, \quad & &   P_2 &\; =\;& \frac{J_{6s}}{ 8 J_{2s}} \;, \\
& P_4' &\; =\;&  \frac{J_4}{\sqrt{-J_{2s} J_{2c}}} \;, \quad & &  P_5' &\; =\;&  
\frac{J_5}{2\sqrt{-J_{2s} J_{2c}}}   \;, \\
& P_6'  &\; =\;&  \frac{J_7}{2 \sqrt{-J_{2s} J_{2c}}} \;, \quad & &  
P_8' &\; =\;&  \frac{J_8}{2\sqrt{-J_{2s} J_{2c}} }  \;, \\
& A_{\rm FB} &\; =\;& \frac{J_{6s} + J_{6c}/2}{d\Gamma/dq^2}     
\;, \quad & &  F_L &\; =\;&   \frac{J_{1c} - 1/3 J_{2c}}{d\Gamma/dq^2} \;.
\end{alignat}
The $J_i$-functions appear in the differential distribution 
\begin{eqnarray}
\nonumber
& &   \frac{8 \pi}{3 \vv \vv_\ell}  \frac{d^4 \Gamma}{d q^2\, d\!\cos\thl\, d\!\cos\thK\, d\phi}  =     (J_{1s} + J_{2s} \cos\!2\thl + J_{6s} \cos\thl) \sin^2\!\thK
  + (J_{1c} + J_{2c} \cos\!2\thl + J_{6c} \cos\thl) \cos^2\!\thK   +
\\ \nonumber
   &  & (J_3 \cos 2\phi + J_9 \sin 2\phi) \sin^2\!\thK \sin^2\!\thl
+  (J_4 \cos\phi + J_8  \sin\phi) \sin 2\thK \sin 2\thl 
+  (J_5 \cos\phi  + J_7 \sin\phi ) \sin 2\thK \sin\thl \, ,
\label{eq:anganal}
\end{eqnarray}
and are of products of two helicity amplitudes.
In \eqref{eq:anganal} $\theta_\ell$ stands for the angle between the $\ell^-$ and
$\bar{B}$ in the $(\ell^+\ell^-)$ centre of mass system (cms), 
$\thK$ the angle between $\bar{B}$ and $K^-$  in the $(K^-\pi^+)$ cms and 
$\phi$ the angle between the two decay planes 
%spanned by the 3-momenta of the
%$(K^-\pi^+)$- and $(\ell^+\ell^-)$-systems
, respectively.  
The variables $\vv$ and $\vv_\ell$ denote the momentum of the $K^*$ meson and the $\ell$ in the rest frame for the $B$-meson and the lepton pair respectively e.g. \cite{HZ13}.

The averages, reflect the fact that the in the experiment the $J_i$'s are fitted for in separate bins. 
For example for $P_5'$ this amounts to,
\begin{equation}
\aver{P_5'}_{[a,b]} =  \frac{\aver{J_5}_{[a,b]} }{2\sqrt{-\aver{J_{2s}}_{[a,b]}  \aver{J_{2c}}_{[a,b]} }} \quad \text{with} \quad  \aver{J_{5}}_{[a,b]} \equiv \int_a^b  dq^2J_{5}(q^2) \;,
\end{equation} 
with all other cases being analogous.
%P3[q2_, t_: pQCD, nb_: 1] := 
%  Module[{}, type = t; etaB = nb; -J9[q2]/(4 J2s[q2])];

\section{Details on fits}
\label{app:fit}

\subsection{BESII-charmonium fit}
\label{app:BESII}

The result of the 16 fit parameters to the BESII-data fit \eqref{eq:Rfit} is given in table \ref{tab:BESfit}.

\begin{table}[h]
\parbox[t]{12cm}{
\begin{center}
\begin{tabular}{|c|c|c|c|c|c|} \hline \hline
    $r$    &         &$\psi(3770)$  &$\psi(4040)$&$\psi(4160)$&$\psi(4415)$\\ \hline
 %       &PDG2004  &3769.9$\pm$2.5&4040$\pm$10 &4159$\pm$20 & 4415$\pm$6 \\
 %       &PDG2006  &3771.1$\pm$2.4&4039$\pm$1.0&4153$\pm$3  & 4421$\pm$4 \\
   %  &CB (Seth) &   -          &4037$\pm$2  &4151$\pm$4  & 4425$\pm$6 \\
 % &BES (Seth)&   -          &4040$\pm$1  &4155$\pm$5  & 4455$\pm$6 \\
 
  $m_r$      &BES\cite{BES08} &  3771.4(18) & 4038.5(46) & 4191.6(60) & 4415.2(75) \\
 (MeV/$c^2$)         &our fit&3771.0(21)& 4036.9(47) & 4190.3(82) & 4416.0(114)
      \\  \hline\hline 
$\Gamma_r$&  BES\cite{BES08}       &25.4(65) &81.2(144) &72.7(151) &73.3(212)  \\
 (MeV)         &our fit&  23.3(51)&76.2(151)&73.5(429)&78.5(571) 
             \\\hline\hline
$\Gamma^{r \to e^+e^-}$& BES\cite{BES08}  & 0.22(5) & 0.83(20) & 0.48(22) &  0.35(12) \\
(keV)  &   our fit   &   0.23(5) & 0.76(15) & 0.73(43) &  0.79(57)\\
              \hline\hline
              % M=3.771004(0.002145) GeV, GammaEE=0.183941(0.038850) keV, Gamma=23.288414(5.079924) MeV, delta=0.000000(0.000000)                                                                                                                
%M=4.036868(0.004706) GeV, GammaEE=0.864613(0.199973) keV, Gamma=76.159049(15.056774) MeV, delta=160.299811(53.980050)                                                                                                            
%M=4.190316(0.008248) GeV, GammaEE=0.560487(0.385704) keV, Gamma=73.468715(42.941222) MeV, delta=336.495417(57.733634)                                                                                                            
%M=4.415973(0.011400) GeV, GammaEE=0.434895(0.303773) keV, Gamma=78.503301(57.102804) MeV, delta=289.589655(65.582563)         
$\delta_r$ (degree)&BES\cite{BES08} &0&133(68)&301(61)&246(86)\\
&our work &0&160(54)&337(57)&290(66)\\

\hline\hline
\end{tabular}
\caption{The resonance parameters for the fit ansatz \eqref{eq:T} 
 The total width $ \Gamma_r = \Gamma_r(m_r)$ in Eq.~\eqref{eq:T}.  The parameter of the background model \eqref{eq:Rcon} fit is $a = 3.04$.
 Note there are some minor differences between the published version \cite{BES08} and the arXiv-version. 
 We have taken the values from the published version. We get $\chi^2/$\dof$ = 1.015$ whereas BESII quotes  $\chi^2/$\dof$= 1.08$.}
\label{table}
\label{tab:BESfit}
\end{center}}
\end{table}

\subsection{Combined BESII and LHCb fit}
\label{app:combined}

The main fit parameters, the ones relevant to the LHCb data, are given in table \ref{tab:fit1} in the main text. 
The remaining fit parameters of the charmonium data are now shown. Their values are very close to the ones given 
in table \ref{tab:BESfit} since (i) the BESII-data has small uncertainties compared to the LHCb data  
(ii) the model \eqref{eq:Rfit} is a good ansatz. For this fit we have employed a $\chi^2$-minimisation on a logarithmic scale 
in order to avoid a systematic downwards shift. Since the LHCb errors essentially scale with the central value there is a benefit 
for the fit to shift downwards since the distance to higher points is larger in terms of uncertainty distance. 
This results in low $\n$-values. For example $\n \simeq 0.5$ for fit-a) which clearly is not sensible. 
Effectively the log-scale fit amounts to replacing 
\begin{equation}
\chi^2 = \frac{(x-y)^2}{\sigma^2}  \quad \to \quad  \chi^2 = \frac{(\ln x-\ln y)^2}{(\sigma/y)^2} \;,
\end{equation}
where  $x$ is the model,  $y$ the experimental data and $\sigma$ its associated  uncertainty respectively. 
These two expressions are identical at leading order in $(x/y-1)$. As can be inferred from the table \ref{tab:fit1} 
the values of $\n$ are now centred around one which seems a more likely outcome.
The $\chi^2/\dof$ of fits b),c) and d) only change minimally whereas  $\chi^2/\dof|_{fit \;a)} = 2.18$ as opposed to $3.59$.  Yet we think at this level this does not have much meaning.

%\section{Anomalous tresholds}
%\label{app:anomalous}
%To be done ... not sure how extensive

\end{document}